\documentclass[a4paper,fleqn]{cas-sc}
\usepackage[numbers]{natbib}
\usepackage[T1]{fontenc}   % 启用 T1 编码（支持欧洲语言字符）
\usepackage[utf8]{inputenc} % 确保正确读取 UTF-8 编码的 .bib 文件
\usepackage{graphicx}%
\usepackage{multirow}%
\usepackage{amsmath,amssymb,amsfonts}%
\usepackage{amsthm}%
\usepackage{mathrsfs}%
\usepackage[title]{appendix}%
\usepackage{xcolor}%
\usepackage{textcomp}%
\usepackage{manyfoot}%
\usepackage{booktabs}%
\usepackage{algorithm}%
\usepackage{algorithmicx}%
\usepackage{algpseudocode}%
\usepackage{listings}%
\usepackage{tabularx}    % 自动调整列宽
\usepackage{booktabs}    % 专业表格线
\usepackage{makecell}    % 列头换行
\usepackage{rotating}    % 横向表格
\usepackage{placeins}    % \FloatBarrier
\usepackage{caption}     % 优化标题间距
\def\tsc#1{\csdef{#1}{\textsc{\lowercase{#1}}\xspace}}
\tsc{WGM}
\tsc{QE}
\tsc{EP}
\tsc{PMS}
\tsc{BEC}
\tsc{DE}
\begin{document}
% \let\WriteBookmarks\relax
% \def\floatpagepagefraction{1}
% \def\textpagefraction{.001}

\shorttitle{PDE Modeling of Norm Dynamics in Multi-Agent Systems}

\shortauthors{Chao Li et~al.}

\title [mode = title]{Modeling Descriptive Norms in Multi-Agent Systems: An Auto-Aggregation PDE Framework with Adaptive Perception Kernels}                      
\tnotemark[1]

\tnotetext[1]{This research is financially supported by The Russian Science Foundation, Agreement 24-11-00272.}

% \tnotetext[2]{The second title footnote which is a longer text matter
%    to fill through the whole text width and overflow into
%    another line in the footnotes area of the first page.}

\author[1]{Chao Li}[orcid=0009-0000-1233-3281]

\cormark[1]

\fnmark[1]

\ead{316325@niuitmo.ru}

% \ead[url]{www.cvr.cc, cvr@sayahna.org}

\credit{Ideas, experiments, code, and the paper writing}

\affiliation[1]{organization={ITMO University},
    city={Saint Petersburg},
    % citysep={}, % Uncomment if no comma needed between city and postcode
    postcode={197101}, 
    country={Russian Federation}}

\author[1]{Ilia Derevitskii}[orcid=0000-0002-8624-5046]
\fnmark[2]
\ead{ilyaderevitskiy@gmail.com}
\credit{Dataset processing}

\author[1]{Sergey Kovalchuk}[orcid=0000-0001-8828-4615]
\fnmark[3]
\ead{kovalchuk@itmo.ru}

\credit{Research guidance}

\cortext[cor1]{Corresponding author}
% \cortext[cor2]{Principal corresponding author}

\fntext[fn1]{Corresponding author. E-mail address:316325@niuitmo.ru}
% \fntext[fn2]{Another author footnote, this is a very long footnote and
%   it should be a really long footnote. But this footnote is not yet
%   sufficiently long enough to make two lines of footnote text.}

% \nonumnote{This note has no numbers. In this work we demonstrate $a_b$
%   the formation Y\_1 of a new type of polariton on the interface
%   between a cuprous oxide slab and a polystyrene micro-sphere placed

\begin{abstract}
This paper presents a PDE-based auto-aggregation model for simulating descriptive norm dynamics in autonomous multi-agent systems, capturing convergence and violation through non-local perception kernels and external potential fields. Extending classical transport equations, the framework represents opinion popularity as a continuous distribution, enabling direct interactions without Bayesian guessing of beliefs. Applied to a real-world COVID-19 dataset from a major medical center, the experimental results demonstrate that: when clinical guidelines serve as a top-down constraint mechanism, it effectively generates convergence of novel descriptive norms consistent with the dataset; in the bottom-up experiment, potential field guidance successfully promotes the system's reconstruction of descriptive norms aligned with the dataset through violation-and-recoupling; whereas fully autonomous interaction leads to the emergence of multi-centric normative structures independent of the dataset.
\end{abstract}

% \includegraphics{figs/grabs.pdf}

% \begin{highlights}
% \item Research highlights item 1
% \item Research highlights item 2
% \item Research highlights item 3
% \end{highlights}

% Keywords

\maketitle

\section{Introduction}

Descriptive norms are an abstract summary of humans' own collective tendencies~\cite{Donaldson1994TowardAU,Goldstein2007UsingSN,Morris2015NormologyII}. They manifest as statistical regularities in group behavior or thought ~\cite{Scharding2023WhenAN,Kalantari2013OnTL}. Descriptive norms are transmitted and shared through interactions, which characterize human communication, behavior, and thought tendencies via dynamic observations and informal information dissemination ~\cite{Kasper2020PracticalRF,Phi2012ExpandingTG}. 

Modeling using autonomous agent systems requires simultaneously representing both this subjective perception and the collective actual tendencies of humans~\cite{chen2024social}. At the same time, descriptive norms, as a dynamic structured opinion, exist in the form of weakly formalized, non-strict natural language expressions. They are more like an individual's speculation and perception of the collective.

To capture this dynamic, abstract descriptive norm, we were inspired by computational methods of PDEs in continuous opinion spaces~\cite{mogilner1999non,keller1970initiation,horstmann20031970,boi2000modeling,hillen2009user} and introduced an auto-aggregation model, representing agent movement in the opinion space through gradient climbing in the opinion space, using the distribution formed by this collective movement to characterize collective descriptive norms.

The driving force for agents' convergence to and violation of descriptive norms in the opinion space comes from the gradient of the opinion popularity equation itself. We extended this perception-kernel-based non-local gradient theory\cite{mogilner1999non,di2013measure,boi2000modeling,sayama2020enhanced} by introducing an external spatial potential field into the transport equation of opinion dynamics.

Our approach no longer uses Markov games and "approximate" rational Bayesian methods that are based on probabilistic guessing of others' beliefs about norms~\cite{Tan2018ThatsML,Cranefield2015ABA,Nichols2016RationalLA,chen2024social}. Human thinking is opaque, and for agents interacting with humans, guessing is necessary, but within autonomous multi-agent systems, agents can directly interact through kernels for non-local perception. The kernel function can be parameterized to model agents' breadth of information gathering and selective attention to dissimilar viewpoints, thus forming auto-convergence or auto-avoidance of the collective distribution, thereby achieving not only convergence to descriptive norms but also violation of norms, thus forming a complete representation of opinion dynamics for the propagation and sharing of descriptive norms.

To avoid completely hypothetical experiments, this study systematically investigates the dissemination and sharing mechanisms of descriptive norms through numerical experiments using mathematical models based on real COVID-19 medical data from large medical centers in major countries. Medical practices are strictly constrained by century-evolved norms, possessing all-scenario rigid characteristics (all operations embedded within the norm framework), significantly differing from other social domains (e.g., in China, physicians with doctoral degrees must complete three years of standardized training); and norms continue to dynamically evolve, rendering the rapidly changing environment of the COVID-19 pandemic an ideal scenario for validating the aforementioned mechanisms~\cite{han2024covid,chen2021impact,xiao2020china}.

\begin{figure}
  \centering
  \includegraphics[width=0.9\columnwidth]{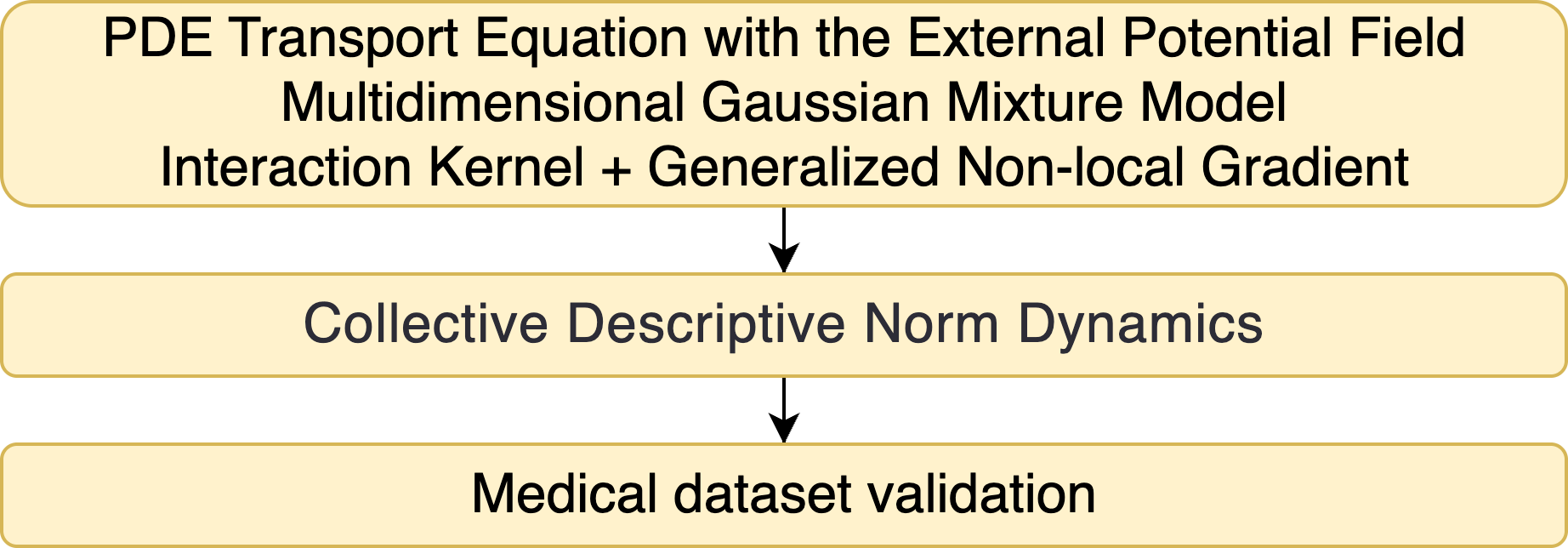} 
  \caption{Validating our descriptive norm model on the medical dataset}
  \label{fig:dynamic_model}
\end{figure}

Integrating this dataset with expert insights, we identified two empirical patterns: (1) National clinical guidelines induce top-down convergence toward new collective descriptive norms; (2) Emerging variants trigger bottom-up practice shifts that violate prior guidelines, generating revised descriptive norms. Our objective is to model these top-down/bottom-up norm dynamics through an integrated agent-based framework, capturing differential propagation patterns via computational inputs—including variant emergence timelines, guideline releases, and practice-fact interdependencies. An abstract diagram of our model can be seen Figure~\ref{fig:dynamic_model}).

%%%%%%%%%%%%%%%%%%%%%%%%%%%%%%%%%%%%%%%%%%%%%%%%%%%%%%%%%%%%%%%%%%%%%%%%

\section{Related work}
The key to constructing a computational model of descriptive norms lies in bridging the multidimensional construction across individual/collective, belief/behavior, and subjective/objective dimensions. The approximate Bayesian approach of "observation-guessing" is not suitable for characterizing the propagation and sharing patterns of descriptive norms as collective descriptive content~\cite{Oldenburg2024LearningAS}. Moreover, many similar models~\cite{axelrod1981evolution,savarimuthu2009norm,vinitsky2023learning,cushman2019punishment,yan2024agent} tend to use hypothetical sanctions/punishments to regulate agents' learning of norms, yet punishments for violating descriptive norms in real society often do not come immediately, even in the medical field. Moreover, such violations may not be based on negative motivations. We also focus on the cognitive level of norm learning, but do not consider punishment as the primary factor~\cite{vinitsky2023learning}. This is because "punishment beams" in games~\cite{leibo2021scalable,agapiou2022melting}, learned strategies, or metanorms (which mandate punishment of violators) are all adopted when agent motivations cannot be controlled. However, in the propagation of descriptive norms, the complex system stratification problem of subjective-objective, individual perception range, and collective actual tendencies is the core. Unlike the views of~\cite{tan2019bayesian,Theriault2019TheSO,Kelly2018SOCIALNA}, we argue that descriptive norms are not simply observations of statistical regularities and are not common sense lacking "oughtness," which is particularly evident in medical datasets. We acknowledge the view that continuous interaction in distributed scenarios serves as real norm perception~\cite{Dingemanse2023BeyondSA,mirski2021conventional}, rather than simply inferring (guessing) norms based on behavior~\cite{shum2019theory,tan2019bayesian}. The so-called enemy/friend distinction and identifying cooperative/competitive objects are not solid assumptions for the generation of descriptive norms~\cite{shum2019theory,galinsky2016friend}. Modeling others' emotional reasoning~\cite{ong2015affective} and inferring others' beliefs and desires from emotional expressions~\cite{wu2018rational} do not align with modeling within complex systems; they are more suited for human-machine collaboration. Social pressure leads to the propagation of "ought" meaning, and the gradient in our model can produce similar effects~\cite{Theriault2019TheSO}. The modeling approach of weighing costs and benefits to decide actions can also be extended to our generalized non-local gradient~\cite{jara2016naive}. Ethical norms can also serve as our dataset selection, simply extending clinical guidelines and practice norms to moral inclinations.

\section{Mathematical Model}

$P(x, t)$ represents the population size holding opinion x at time t. Our auto-aggregation mathematical model ~\cite{keller1970initiation,boi2000modeling,hillen2009user,sayama2020enhanced,horstmann20031970} extends the function $P(x, t)$ describing opinion popularity dynamics to describe the group's violation and conformity to collective descriptive norms. Descriptive norms propagate and are shared within the collective, manifested as changes in opinion popularity.  

The model assumes a constant total population, with distribution changes only due to diffusion and migration: diffusion reflects random opinion fluctuations, while migration represents directed active opinion shifts induced by social influence.  

Specifically, the mathematical model is based on the following assumptions: individuals only perceive information from neighboring regions in opinion space ~\cite{bakshy2015exposure}; individuals not only learn norms but also decide when to comply, thereby adjusting perception range and direction ~\cite{kwon2023not,levine2020logic}; different individuals have subjective perceptions of "what constitutes normal behavior" ~\cite{chen2024social}; the direction of group opinion movement forms a dynamic interaction between norm perception and group behavior. 

We extend the classical transport equation framework \cite{hillen2009user,di2013measure}, and the mathematical description of P(x,t) is: 

\begin{equation}\label{eq:1}
\frac{\partial P}{\partial t} = d \nabla^2 P - \nabla \cdot \left[ P \left( c  G(P) - \nabla V(x) \right) \right],
\end{equation}

where $d \nabla^2 P$ is the diffusion term and $- \nabla \cdot \left[ P \left( c  G(P) - \nabla V(x) \right) \right]$ is the migration term. $V(x) = k \cdot (x - x_{\text{target}})^2$ is the external potential field function, following the physical principle that force is the negative gradient of potential energy, $k$ is the potential strength parameter, and $x_{\text{target}}$ is the target position. In the absence of an external force field with spatially dependent guidance (such as macroscopic norm regulation), the migration term automatically reduces to \(-c \nabla \cdot (P G(P))\). $G(P)$ is the perceived gradient of popularity distribution, defined as:

\begin{equation}\label{eq:2}
G(P) = \int_{-\infty}^{\infty} P(x + y, t) g(y) dy,
\end{equation}

\begin{equation}\label{eq:3}
g(y) = \frac{1}{2\mu} \cdot \frac{1}{\sqrt{2\pi}\sigma} \left( e^{-\frac{1}{2}\left(\frac{y-\mu}{\sigma}\right)^2} - e^{-\frac{1}{2}\left(\frac{y+\mu}{\sigma}\right)^2} \right).
\end{equation}

The external potential field is conditionally activated based on the number of opinion clusters present in the system, allowing for dynamic control of opinion aggregation behavior while maintaining the fundamental structure of the original transport equation model. 

The opinion movement patterns generated by the PDE-based opinion dynamics can be formally modeled as collective descriptive norms, where the emergent spatial-temporal configurations in Equation~\ref{eq:1} directly manifest the statistical properties of population-level consensus. 

Given the assumption that individuals possess subjective perceptions of "what constitutes normal behavior"~\cite{chen2024social}, we can characterize descriptive norms using multidimensional Gaussian mixture models that integrate both subjective and objective perceptions~\cite{chen2024social,li2025multi}. The objective collective norm (OBJ), defined in Equation~\ref{eq:obj}, quantifies population-level consensus through a weighted mixture of $K$ Gaussian groups. Crucially, the subjective individual norm perception (SINP) in Equation~\ref{eq:SINP} dynamically evolves via decentralized interactions, forming an adaptively weighted Gaussian mixture that reflects real-time inferences of collective norms. 
\begin{eqnarray}\label{eq:obj}
\text{OBJ} & = & \sum_{i=1}^{K} w_i \cdot g(x|\mu_{\text{group}}^i, \sigma_{\text{group}})
\end{eqnarray}
% %
% \begin{eqnarray}\label{eq:Tendcy}
% \text{IndT}_j & \sim & \mathcal{N}(\mu_{\text{IndT}}^j, \sigma_{\text{indiv}})
% \end{eqnarray}
% %
\begin{eqnarray}\label{eq:SINP}
\text{SINP}_j & = & \sum_{i=1}^{K} w_{\text{sub}}^i \cdot g\left(x|\mu_{\text{sub}}^i, \sigma_{\text{sub}}\right)
\end{eqnarray} 

The OBJ can be derived via dataset analysis and pattern recognition modeling, with distinct acquisition approaches applicable to different datasets. At time $t$, the collective distribution formed by $p(x,t)$ across all $x$ constitutes the objective collective tendency or objective collective descriptive norm. 

We model each $x_j$ at distinct positions of $P(x,t)$ as autonomous agents endowed with their respective $\text{SINP}_j$. The perceptual kernel function in Equation~\ref{eq:3} captures bilateral popularity disparities through a composite of two internal Gaussian distributions, enabling agents to assimilate localized information and formulate subjective norm perceptions. $\text{SINP}_j$ at each spatial location constructs a Gaussian Mixture Model during initialization by mapping peak positions, squared half-widths, and normalized heights from the initial distribution within a $\sigma_{\text{init}}$-$|\mu_{\text{init}}|$-determined adaptive window to Gaussian means, variances, and weights; $\text{SINP}_j$  dynamically updates this model using peak characteristics extracted via kernel density estimation with bandwidth-controlled smoothing of locally sampled opinion data, maintaining the parameter mapping principle while adapting to evolving population dynamics. The initialization and update procedures for $\text{SINP}_j$ are predominantly implemented through algorithmically designed code based on mathematical formalizations, which admit multiple implementation variants rather than being intrinsically defined by mathematical formulations. Our customized code architecture may be consulted in Appendix~\ref{appendix:SINP}.

The generalized nonlocal gradient is a differential operator extended via kernel function. Equation~\ref{eq:2} demonstrates that the perceptual gradient $G(P)$ represents the cross-correlation quantity between perception and kernel $g$ – a classical interaction kernel approach in applied mathematics and physics~\cite{boi2000modeling,di2013measure,mogilner1999non,sayama2020enhanced}. This perceptual kernel governs how individuals weight neighboring opinion popularity when evaluating gradients. 

Where $\mu > 0$, the migration process simplifies to climbing behavior along the gradient of $P(x,t)$, manifesting as auto-aggregation that converges toward collective descriptive norms. When $\mu < 0$, the kernel polarity reverses ($g_{\mu}(y) = -g_{-\mu}(y)$), inducing auto-avoidance behavior that violates collective descriptive norms. 

Increasing $\sigma$ signifies enhanced information gathering capacity (broader opinion coverage), while increasing $\mu$ elevates attention to dissenting views (heightened focus on distant opinions). When $\mu \to 0^+$ and $\sigma \to 0^+$ (i.e., $g$ degenerates to signed Dirac $\delta$-functions about the origin), $G(P)$ converges to the conventional derivative $\partial P/\partial x$, establishing $G(P)$ as a mathematically rigorous non-local generalization of spatial derivatives~\cite{di2013measure,mogilner1999non,sayama2020enhanced}. This mathematical representation establishes a rigorous correspondence between microscopic agent interactions and macroscopic normative structures, demonstrating how the solution trajectories of the opinion dynamics PDE fundamentally encode the evolving descriptive norms through their characteristic clustering patterns and stability properties.

\section{Numerical Experiments Setup}
\subsection{Data and Problem Description}
This computable model extends to multiple datasets. Using real-world COVID-19 data from a major global medical center (6,188 observational records, 1,992 unique cases indexed by identifiers like \texttt{"GACAk+Q"}), treatment data comprises 33 controlled columns (medications/procedures), 14 monitored columns (status changes), and 7 metadata columns, with prefixes in patient status features; full headers in Appendix \ref{appendix:Dataset Description}.  

\begin{figure}
    \centering
    \includegraphics[width=\textwidth]{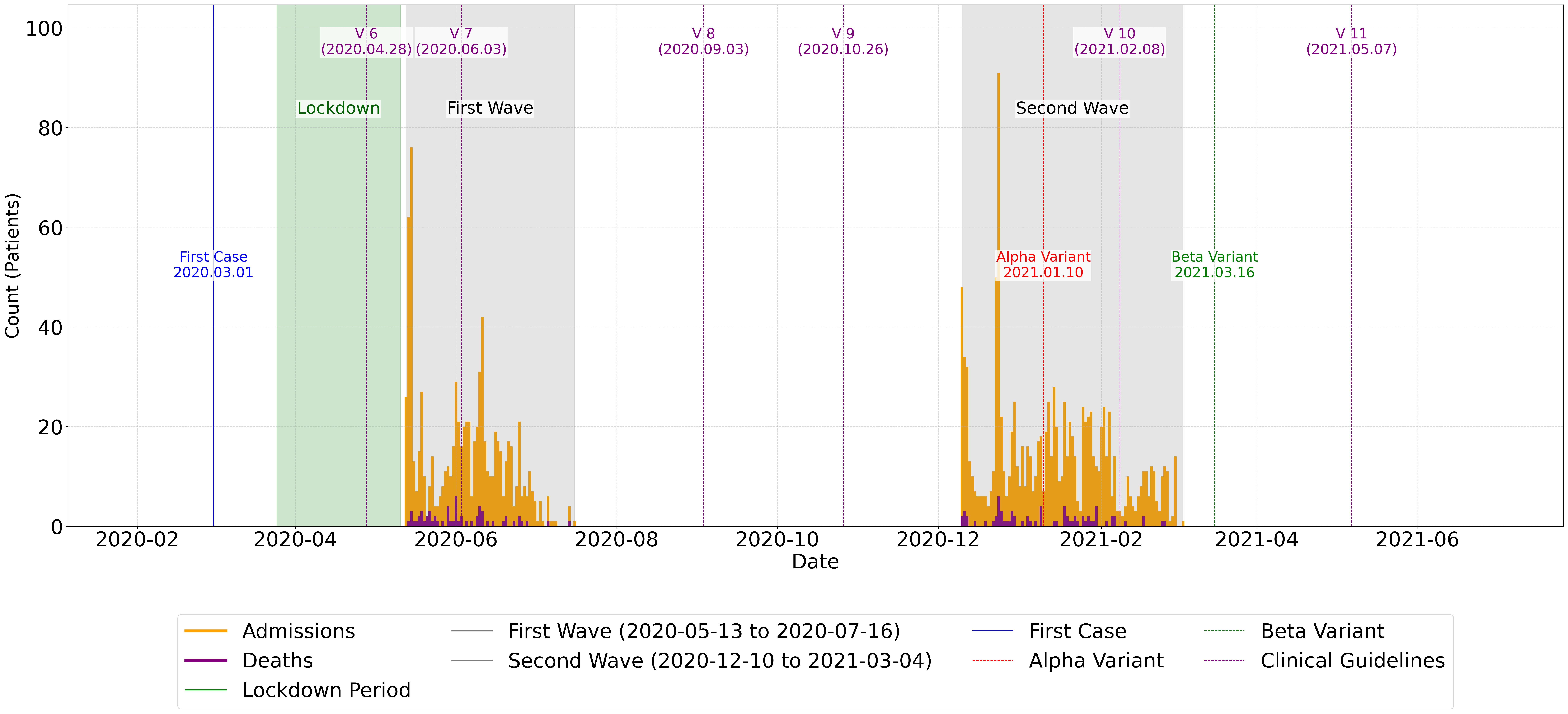} % 使用全栏宽度
    \caption{Key event timeline: COVID-19 guideline releases (6th–11th editions); pandemic waves (Wave 1 start: 13 May 2020; Wave 2 end: 4 March 2021); first case detection; lockdown; Alpha/Beta variant identification at medical center.}
    \label{fig:covid_timeline}
\end{figure}

The dataset directly incorporates clinical records from patients treated at this medical center, spanning the period from the onset of the country's first pandemic wave on 13 May 2020 until the effective conclusion of the second wave on 4 March 2021. Additionally, it includes the official release dates at this medical center for national COVID-19 clinical guidelines, ranging from the 6th to the 11th edition. Key epidemiological event dates recorded at this center are also incorporated, namely: the detection date of the first confirmed COVID-19 case, the implementation date of lockdown measures, and the dates of first detection for the Alpha and Beta variants. Details are illustrated in Fig.~\ref{fig:covid_timeline}.

Based on the three key events at the medical center—receiving the 7th edition guideline, first detection of the Alpha variant, and receiving the 10th edition guideline—we divided the two pandemic waves into 5 periods based on admission time, numbered in ascending numerical order. Combining the dataset with expert opinions, we manually identified the following two empirical patterns in the data from the five periods: 1. After the release of national macro-level clinical guidelines, top-down direct influence on micro-level practice tendency changes leads to the generation of new collective descriptive norms, manifesting a simple "convergence" relationship with the new norm pattern; 2. Changes in medical reality due to new variants prompt micro-level medical control project practice tendency changes, generating new descriptive norms, and forming a "violation" relationship with previous clinical guidelines.

Our experimental objective is to quantitatively characterize the top-down and bottom-up dynamic evolution of "collective descriptive norms" embedded in both patterns through an integrated mathematical framework combining an autonomous multi-agent partial differential diffusion-migration model with a multidimensional Gaussian mixture model. This approach enables our mathematical model to capture the differential patterns of norm propagation and sharing observed through manual inspection. To achieve this objective, computational modeling of changes in medical practice tendencies, shifts in medical facts, impact of changes in medical facts on medical practice tendencies, temporal occurrences of new variant emergences and clinical guideline releases across phases using various computational methods including machine learning are implemented as model inputs.
Comprehensive parameter specifications are summarized in Table \ref{tab:parameters_descriptive}.

\begin{table}[ht]
	\caption{Parameter summary}
	\label{tab:parameters_descriptive}
	\begin{tabular}{cccc}\toprule
		Kernel $\sigma$ & Kernel $\mu$ & Potential $k$ & Diffusion $d$ \\
		Migration $c$ & Target $x_{\text{target}}$ & Gradient $G$ & Kernel $g$ \\
		\text{OBJ} & \text{SINP} & GMMs $K$ & Density $P_h$ \\
		Frequency $\omega$ & Phase $\phi$ & Amplitude $\Delta P$ & Domain $L$ \\
		Resolution $N_x$ & $\Delta\mu$ & $\Delta\sigma$ & Potential $V(x)$ \\ \bottomrule
	\end{tabular}
\end{table} 

\subsection{Problem Setup}

To complete the experimental objective, our system dynamics specification for norm evolution includes theoretical validation of the extended PDE framework through linear stability analysis and modeling the medical dynamics from datasets that we need to input into the model.

\subsubsection{Linear Stability Analysis with The Potential Field}

The study of descriptive norm (formation and evolution of 'opinion islands') evolving in space over time is grounded in the numerical condition analysis of linear perturbations. As we extended the migration term of the transport equation, we have rederived the partial differential equations here (see Appendix~\ref{appendix:Details of linear stability analysis} for details). We still begin the analysis by replacing the spatiotemporal function $P(x, t)$ with a constant homogeneous population level $P_h$ plus a sinusoidal spatial perturbation with time-varying small amplitude $\Delta P(t)$~\cite{sayama2015introduction,hillen2009user,di2013measure}, specifically:

\begin{equation}
P(x,t) \rightarrow P_h + \Delta P(t) \sin(\omega x + \phi).
\label{eq:6}
\end{equation}

Through substitution, Equation~\ref{eq:1} can be approximated as the following equation for $\Delta P$:

\begin{equation}
\begin{aligned}
\frac{d \Delta P}{dt} = \bigg(-d \omega^2 & + c \omega P_h \int_{-\infty}^{\infty} \sin(\omega y) g(y)  dy \\
& + 2k \bigg) \Delta P.
\end{aligned}
\label{eq:7}
\end{equation}

By defining \( Q(\omega) = \int_{-\infty}^{\infty} \frac{\sin(\omega y)}{\omega} g(y)  dy \), where the range of \( Q(\omega) \) is strictly confined to the interval \([-1, 1]\), the condition for pattern formation is that for \( \omega > 0 \):

\begin{equation}
Q(\omega) > \frac{1}{c P_h} \left(d - \frac{2k}{\omega^2}\right).
\label{eq:8}
\end{equation}

When Equation~\ref{eq:8} holds, particularly as $Q(\omega) \to 1$ for $\omega \to 0$ yielding the condition $cP_h > \left(d - \frac{2k}{\omega^2}\right)$, the conclusion that the homogeneous population distribution becomes unstable and forms heterogeneous patterns in the opinion space still holds. See Appendix Equation~\ref{eq:A10} for details.

\subsubsection{Computational Modeling of Dynamic Medical Factors}

\paragraph{Continuous propensity field modeling} 
We model the 33 binary treatment features (0/1 values) across five temporal periods as five 33-dimensional Gaussian Mixture Models (GMMs). For each of the five clinical periods, we transformed 33 binary control features (including \texttt{\_stat\_control} and \texttt{\_dinam\_control} variables) into a continuous propensity field~\cite{huber2020direct,brown2021propensity,austin2019assessing}. For patient $i$ and feature $j$, we computed the weighted activation ratio $a_{ij} = \frac{1}{L_i}\sum_{t=1}^{T_i}x_{ij}(t)\cdot(1-0.5\frac{t-1}{T_i-1})$ where $L_i$ is treatment duration, followed by sigmoid smoothing $p_{ij} = 1/(1+e^{-10(a_{ij}-0.5)})$ to obtain the $N \times 33$ propensity matrix $\mathbf{P}$. We then fitted Gaussian Mixture Models using EM algorithm with BIC criterion selecting optimal components $K^*$, and transformed parameters to physical space via mean mapping $\boldsymbol{\mu}_k^{\text{orig}} = \boldsymbol{\mu}_s + \boldsymbol{\sigma}_s \odot \boldsymbol{\mu}_k$ and delta-method covariance approximation $\boldsymbol{\Sigma}_k^{\text{orig}} = \text{diag}((\boldsymbol{\sigma}_s \odot \sqrt{\text{diag}(\boldsymbol{\Sigma}_k)})^2)$, preserving clinical interpretability in $[0,1]^D$ space where $D=33$ represents the dimensionality of the medical decision space. The resulting GMM means are therefore 33-dimensional vectors corresponding to each clinical decision pattern.

Across 5 clinical periods (153–557 patients), BIC-selected GMM components (3–5) reveal distinct decision patterns: Period 2 has 5, others stabilize at 3. Each component = prototypical strategy: $\boldsymbol{\mu}$ = mean feature intensity; covariance = modality co-occurrence. Full stats in Appendix Table~\ref{tab:gmm_results}; top 10 of 33D $\boldsymbol{\mu}$ for P2–P4 = most frequent interventions, in Appendix~\ref{tab:top_features}

\paragraph{Statistical Analysis of Medical Fact Differences} 
Here we employed robust methodologies to compare medical fact distributions between distinct pandemic phases. Data preprocessing differentiated static features (mode imputation for missing values with case-level deduplication) from dynamic features (linear interpolation with case-level median aggregation). Statistical inference utilized the nonparametric Mann-Whitney U test to compare group medians while accommodating non-normal distributions. Effect size quantification implemented Cliff's Delta ($\delta \in [-1,1]$) with directional interpretation ($\delta>0$: Period File 1 > Period File 2) .Multiple testing correction applied Benjamini-Hochberg false discovery rate (FDR) control to maintain $\alpha=0.05$ family-wise error rate, reporting both raw and adjusted p-values. This integrated approach enabled statistically rigorous comparisons between key pandemic phases: Period1  vs Period2 (Appendix Table~\ref{tab:period1_2_comparison})  and Period3  vs Period4 (Appendix Table~\ref{tab:period3_4_comparison}). 

\paragraph{Temporal Causal Interplay Between Medical Facts and Control columns} 

Our data constitutes independent cross-sectional observations per period rather than panel data. Constrained by the sparse data structure (mean 3.22 observations per patient), we prioritize identifying statistically significant correlation patterns between medical facts and control variables over establishing strict causal mechanisms. After rigorous comparative evaluation revealed that cross-sectional DID (OLS with Cluster-Robust SE), Hidden Markov Models (HMM) augmented with change-point detection, and Bayesian Structural Time Series (BSTS) yielded a substantial proportion of non-convergent estimates, Double Machine Learning was selected as the optimal methodology for deriving statistical relationships aligned with domain-expert medical knowledge.

The analytical workflow initiated with data preprocessing: merging $pre-/post-intervention$ datasets using Pandas, imputing missing values via medically grounded zero substitutions, converting admission dates to pandemic-day offsets, and numerically encoding process stages while discarding $t_point$ variables. Automated variable classification segregated $X$ medical facts as causal drivers and $Y$ control columns as outcome variables through naming conventions, supplemented by period dummies. Within the DoubleML framework implementing partial linear regression, we explicitly modeled medical facts as treatment variables ($D$) and control columns as outcome variables ($Y$), employing 5-fold cross-fitting with random forest learners (100 trees, $max_depth$=5) for first-stage predictions. This generated orthogonalized residuals following $Y - \hat{E}[Y|X] = \theta \cdot (D - \hat{E}[D|X]) + \epsilon$, enabling second-stage effect estimation via partialling-out where $\hat{\theta}$ quantifies how changes in medical facts causally influence control outcomes. Causal effects $\hat{\theta}$ were quantified with standard errors $SE(\hat{\theta})$ and p-values derived from $t = \hat{\theta}/SE$. For multiplicity control, Benjamini-Hochberg FDR correction ($q_i = m \cdot p_i / \text{rank}(p_i)$) was applied within control-variable groups, with p-values $<10^{-300}$ stabilized at $10^{-300}$. Diagnostic validation included Jarque-Bera residual normality testing ($JB = n/6(S^2 + (K-3)^2/4)$) and feature importance analysis via Gini impurity rankings. All 240 rows of statistical causal relationship calculations were preserved, with the first 20 rows provided in the Appendix Table~\ref{tab:facts_effec_control} for reference. 

\section{Numerical Experiments}
The baseline parameters are fixed as domain size $L=20.0$, spatial resolution $N_x=1000$, initial density profile $P_0(x)=1.0 + 0.01\sin(0.2x)$, diffusion coefficient $d=0.2$, and migration coefficient $c=1.0$. These settings remain identical across all three numerical experiments to ensure consistent evaluation of the adaptive perception mechanism.

\subsection{Top-Down Convergence from Uniformity}
Direct extension of our five period-specific 33-dimensional GMMs to a PDE framework is infeasible due to the curse of dimensionality: discretizing the continuous propensity space from 1D to 33D would require $1000^{33}$ grid points, rendering computation intractable. To circumvent this, we adopt a dimensionality-reduction strategy where each feature dimension $d \in \{1,\dots,33\}$ is projected onto all GMM components, realizing the objective collective norm (OBJ) through 33 univariate target distributions $\{\mathcal{T}_d\}_{d=1}^{33}$. Each $\mathcal{T}_d$ defines the fitting objective for agents across all positions in the propensity space $\mathit{P}(x,t)$, such that agent behavioral trajectories evolve to match the distribution $\mathcal{T}_d$ corresponding to dimension $d$. This target-driven mechanism forms the foundational premise of our first experiment.

Modeling of Phase 2’s propensity field shows hydroxychloroquine, azithromycin, and chloroquine were wrongly widespread in Wave 1 here, per complex factors~\cite{yazdany2020use,saag2020misguided,dickinson2024exploring}. Post-Wave 1, v8/v9 guidelines flagged insufficient efficacy and cardiac risks for HCQ/CQ, restricting use. Thus, their propensity ranks fell sharply: HCQ (1→28), CQ (8→29), AZI (4→27; co-administered with HCQ) — see Appendix Table~\ref{tab:rank_change_p2_p3}.

\begin{figure}
\centering
\includegraphics[width=0.9\textwidth]{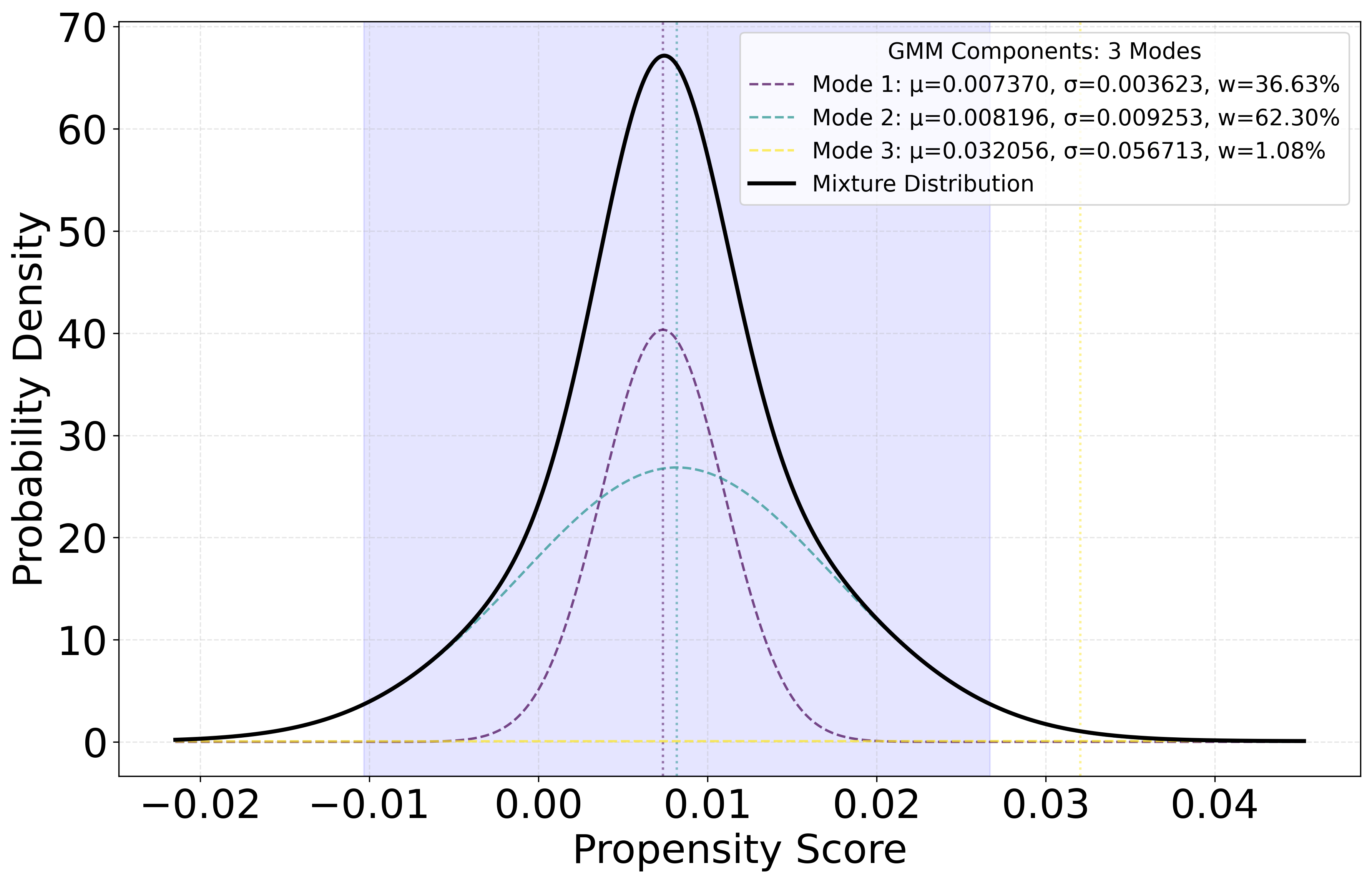}
\caption{
Projection of three components from 33-dimensional Gaussian Mixture Models fitted to Period 3 files for the azithromycin control feature(target OBJ-GMM).}
\label{fig:azithromycin_gmm}

\end{figure}

In this context, we present the descriptive norm pattern for azithromycin control measures learned under v8/v9 guideline guidance, simulated from a uniform initial opinion distribution within the system without using an external potential field term (Fig.~\ref{fig:azithromycin_gmm}).

The Subjective Norm Perception (SINP) model at each spatial location $x_i$ is initialized based on the local population density distribution, where a window radius $r = 5\max(\sigma_{\text{init}},|\mu_{\text{init}}|)$ defines the interval $[x_i-r, x_i+r]$ for detecting local peaks of $P(x)$. Given $K$ detected peaks with positions $\{m_k\}$ and heights $\{h_k\}$, the Gaussian Mixture Model (GMM) parameters are initialized as $\pi_k = h_k/\sum_j h_j$, $\mu_k = m_k$, and $\sigma_k^2 = \max((\text{FWHM}_k/2)^2, 10^{-3})$, where $\text{FWHM}_k$ denotes the full width at half maximum of the $k$-th peak. During simulation, SINP is dynamically updated through weighted sampling: data points are sampled with weights proportional to $P(x)$, and GMM parameters are re-estimated to adaptively capture evolving group norms.

The population density-weighted average Wasserstein distance is computed as $d_{\text{avg}} = \frac{\sum_{i} P(x_i) \cdot d_W(\text{SINP}_i, \text{target})}{\sum_{i} P(x_i)}$, where $d_W$ denotes the Wasserstein distance between the SINP model at location $x_i$ and the target GMM, approximated as $d_W \approx (1/N)\sum_{j=1}^N |x_{(j)} - y_{(j)}|$ with $\{x_{(j)}\}_{j=1}^N$ and $\{y_{(j)}\}_{j=1}^N$ being ordered samples from the respective distributions. Perception kernel parameters are updated according to $\sigma \leftarrow \text{clip}(\sigma + \eta \cdot d_{\text{avg}} \cdot \mathbb{I}(d_{\text{avg}} > \tau), \sigma_{\min}, \sigma_{\max})$ and $\mu \leftarrow \text{clip}(\mu + \eta \cdot d_{\text{avg}} \cdot \mathbb{I}(d_{\text{avg}} > \tau), \mu_{\min}, \mu_{\max})$, where $\eta$ is the learning rate, $\tau$ is the convergence threshold, $\mathbb{I}(\cdot)$ is the indicator function, and $\text{clip}(\cdot)$ enforces boundary constraints, with updates ceasing when $d_{\text{avg}} < \tau$ indicating convergence to the target distribution.

\begin{figure}
\centering
\includegraphics[width=0.9\textwidth]{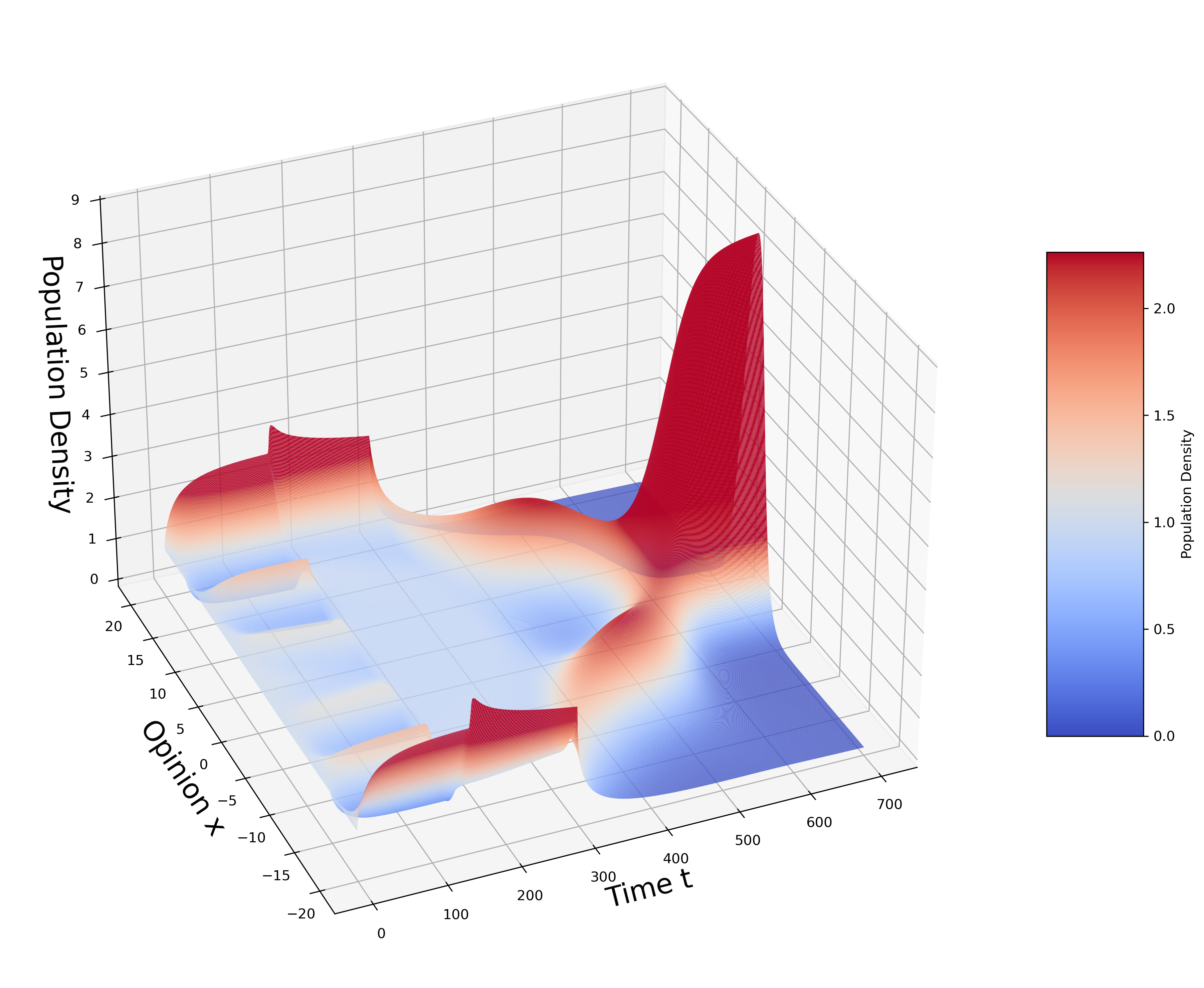
}
\caption{3D surface plot of spatiotemporal population density evolution in opinion space for azithromycin control; $L=20$, $N_x=1000$. }
\label{fig:azithromycin_evolution_3d}

\end{figure}

\begin{figure}
    \centering
    \includegraphics[width=0.9\textwidth]{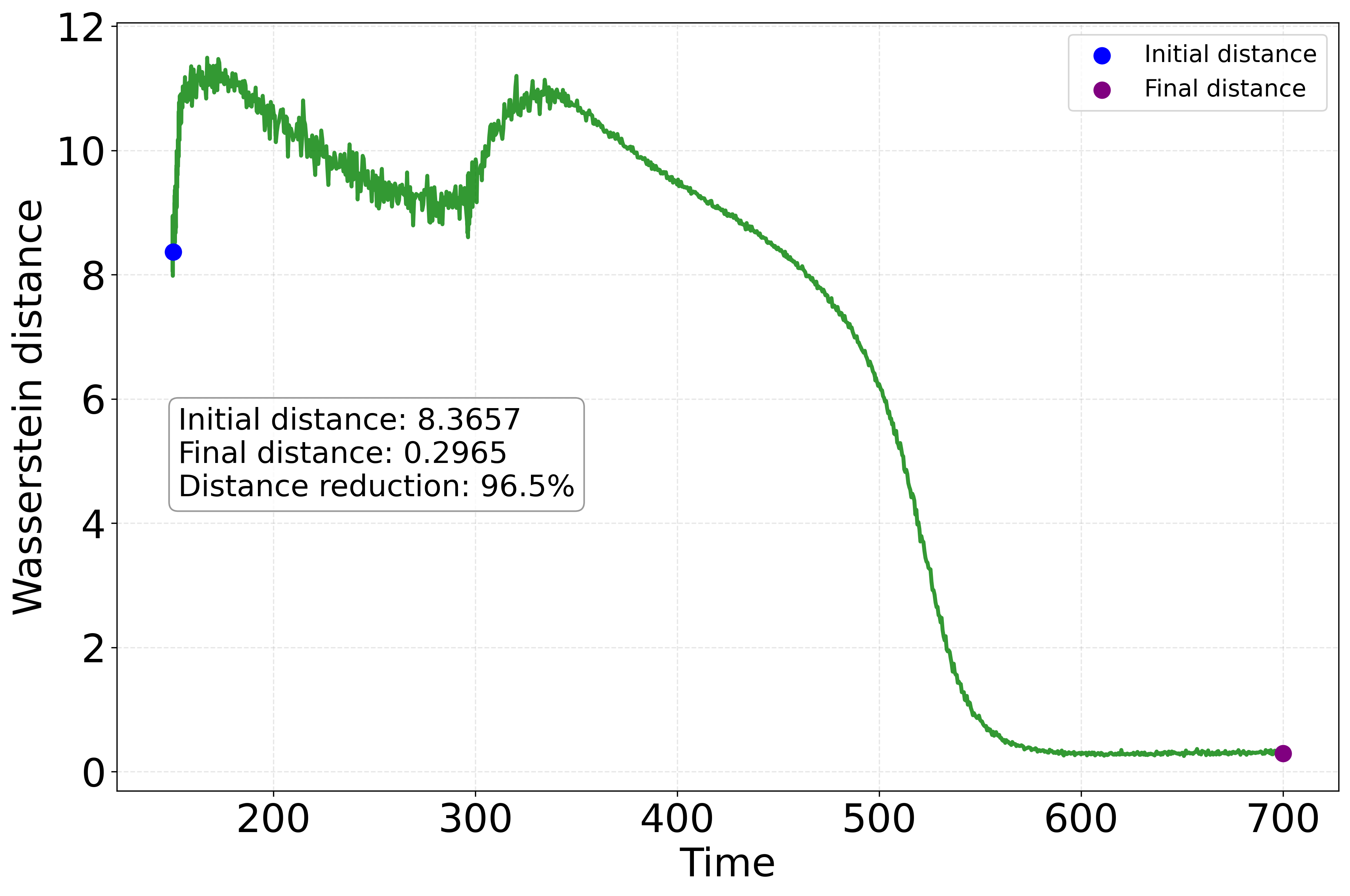}
    \caption{Temporal evolution of Wasserstein distance between simulated population distribution and target GMM for azithromycin control feature}
    \label{fig:azithromycin_distance_evolution}
\end{figure} 

As shown in Figure~\ref{fig:azithromycin_evolution_3d}, we maintain fixed parameter values of $\sigma = 0.1$ and $\mu = -5$ for the first 150 time units. Adaptive updates of the Subjective Norm Perception (SINP) at each spatial location $x_i$ and corresponding kernel parameters $\sigma$ and $\mu$ are activated thereafter. Figure~\ref{fig:azithromycin_distance_evolution} presents the population density-weighted average Wasserstein distance between the descriptive norm (i.e., the opinion distribution composed of all positions in the opinion space $p(x,t)$) and the target objective distribution (azithromycin GMM, cf.~Figure~\ref{fig:azithromycin_distance_evolution}) throughout the adaptive update process up to $t = 700$ time units.

\subsection{Potential Field-Guided Norm Restructuring via Violation Dynamics}

This experiment, based on Experiment 1, employs a bottom-up approach to adjust medical control items according to changes in medical facts, generating new collective descriptive norms. The direct cause of the change in medical facts was the first detection of the Alpha variant at this medical center during the second wave of the pandemic on January 10, 2021. Prior to the release of the 10th version of the national clinical guidelines, the time before and after the emergence of the Alpha variant was divided into two periods (period 3 and period 4), which were modeled as 33-dimensional Gaussian Mixture Models (GMMs). 

The Alpha variant led to increased usage tendency of intermediate-acting corticosteroids Prednisolone (rank 26 to 10) and Methylprednisolone (rank 19 to 13) in Period 4 compared to Period 3. However, it caused decreased usage of the long-acting potent corticosteroid Dexamethasone (rank 4 to 5), which was recommended in versions 8/9 guidelines. Correspondingly, the anticoagulant Enoxaparin Sodium, which was required to be used with corticosteroids according to versions 8/9 clinical guidelines, also decreased (rank rank 16 to 25) following the emergence of the Alpha variant as shown in Appendix~\ref{tab:rank_change_p3_p4}. This is unrelated to disease progression, as Period 4 compared to Period 2 (the latter half of the first wave), also exhibits significant elevation of Prednisolone (rank 32 to 10) as demonstrated in Appendix~\ref{tab:rank_change_p2_p4}.

To capture the practice of continuously adjusting at the micro level and violating the requirements of previous versions of clinical guidelines due to medical fact changes caused by the new Alpha variant, and forming a new collective descriptive norm, we used the following approach for Experiment 2. Here, the medium-potency glucocorticoid Prednisolone is used as the experimental subject.

Based on causal inference analysis using Double Machine Learning methods, Prednisolone as a control variable is statistically significantly influenced by 12 medical Driving Facts. Through the non-parametric Mann-Whitney U test described in Section 4.22, we quantified the intrinsic change intensity (effect\_size\_value) of each medical fact between period 3 and period 4 following the Alpha variant outbreak. Using effect\_size\_value as the first-level weight to characterize the intrinsic dynamic strength of medical facts, and the Causal Effect obtained from statistical causal inference as the second-level weight to reflect its directional and intensity impact on the target variable, we define the doubly-weighted Causal Impact Intensity through their product. By algebraically summing the Causal Impact Intensity of the 12 Driving Facts, we obtained the net causal effect for the Prednisolone control variable.

\begin{figure}
\centering
\includegraphics[width=0.9\textwidth]{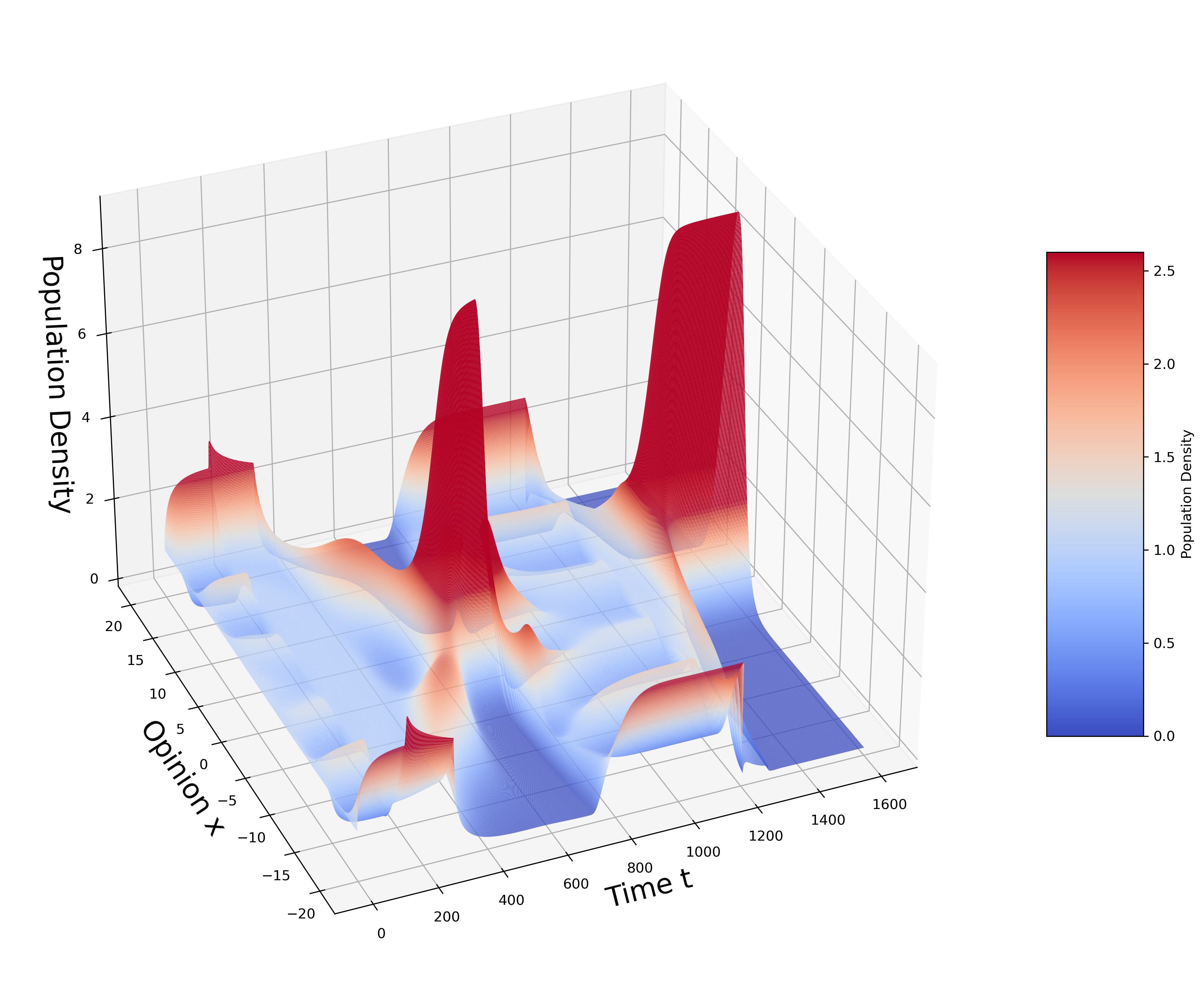}
\caption{Spatiotemporal evolution of population density distribution in opinion space for the Prednisolone control feature.}
\label{fig:prednisolone_evolution_3d_causal_potential}

\end{figure}

We assume that changes in medical fact data serve as external inputs to the agents, directly providing the net causal effects delta\_mu and delta\_sigma. The system quantifies the spatial heterogeneity of opinion distribution by computing the L2 norm of the perceived gradient $G$ (gradient\_strength $= \sqrt{\sum G^2 \cdot dx}$), and maps it to an adaptive factor using the tanh function: adaptation\_factor $= 0.5 \times (1 + \tanh(\text{gradient\_strength} - 0.5))$, making the parameter update rate proportional to the degree of spatial divergence---accelerating adjustments when medical practice opinions are highly divergent and slowing when converging. Simultaneously, the system implements a temporal scaling calibration strategy, where time\_amplification $= 1 + 0.01 \times dt$ achieves a doubling of effect every 100 time units, effectively resolving cumulative effect distortion caused by irregular update intervals. Ultimately, the parameter update formula param\_updater.mu $+$= delta\_mu $\times$ adaptation\_factor $\times$ $dt$ $\times$ time\_amplification (with identical treatment for sigma) organically integrates the three dimensions of net causal effect, spatial heterogeneity, and temporal dynamics.

\begin{figure}
    \centering
    \includegraphics[width=0.9\textwidth]{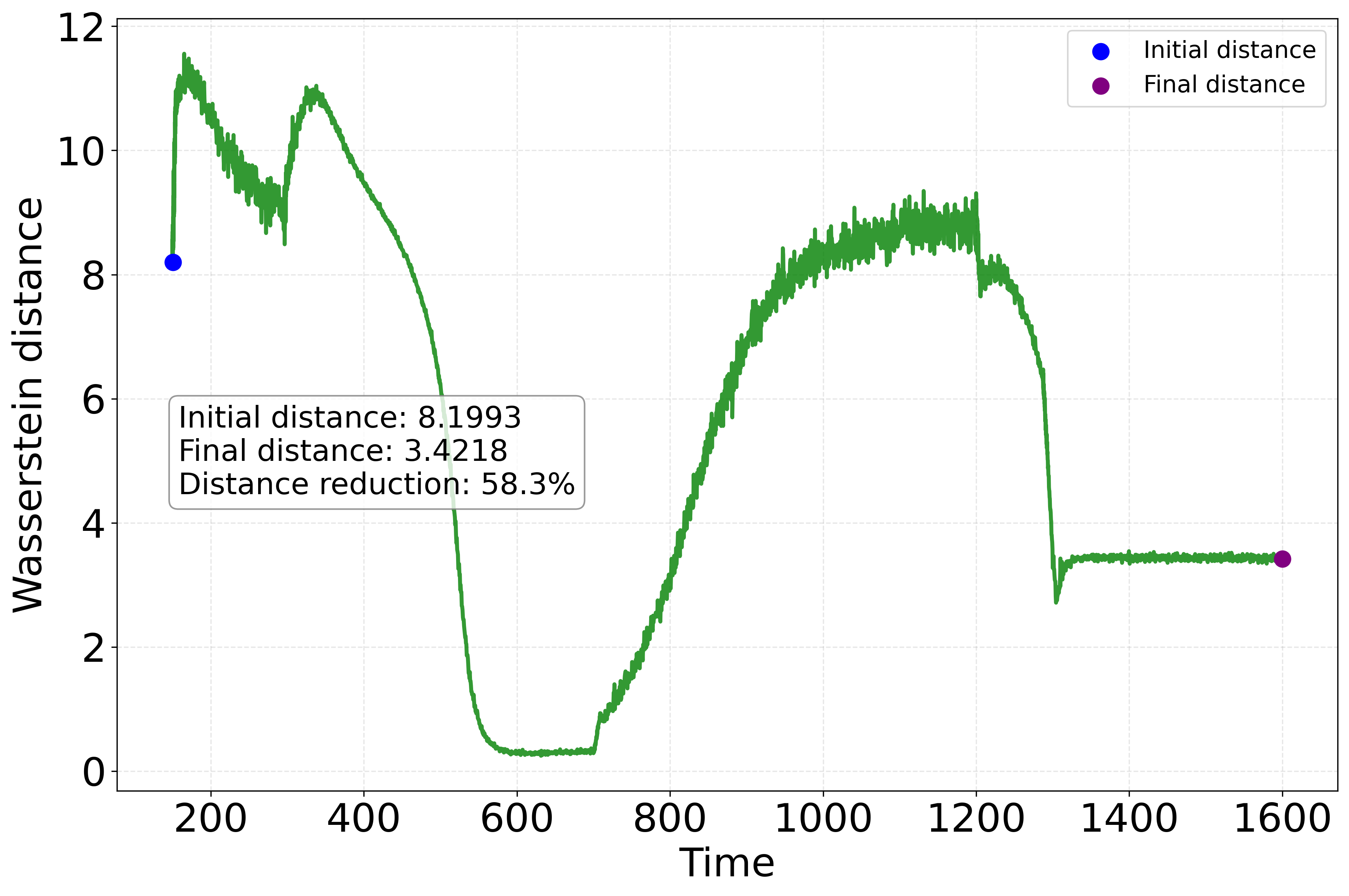}
    \caption{Weighted Wasserstein distance evolution between simulated and target distributions for Prednisolone control: period 3 final distance 0.312819 at $t=700$; period 4 final distance 3.421837 at $t=1600$.}
    \label{fig:prednisolone_distance_evolution}
\end{figure}

This study first employs the method from Experiment 1 to fit the descriptive norm GMMs for Prednisolone in period 3 (final Wasserstein distance of 0.312819 at $t=700.0$). Subsequently, it updates the kernel function parameters based on the causal net effect from Experiment 2. While eliminating the target GMMs setting, the spatial potential field is activated, assuming that despite the initial lack of consensus among healthcare workers when facing the Alpha variant, their long-term professional training experience still generates collective attraction toward a specific position in the opinion space for Prednisolone practice tendencies. The complete fitting 3D visualization and distance evolution are shown in Figures~\ref{fig:prednisolone_evolution_3d_causal_potential} and~\ref{fig:prednisolone_distance_evolution}.

\subsection{Bottom-Up Emergence without Target and Potential Field}
\begin{figure}
    \centering
    \includegraphics[width=0.9\textwidth]{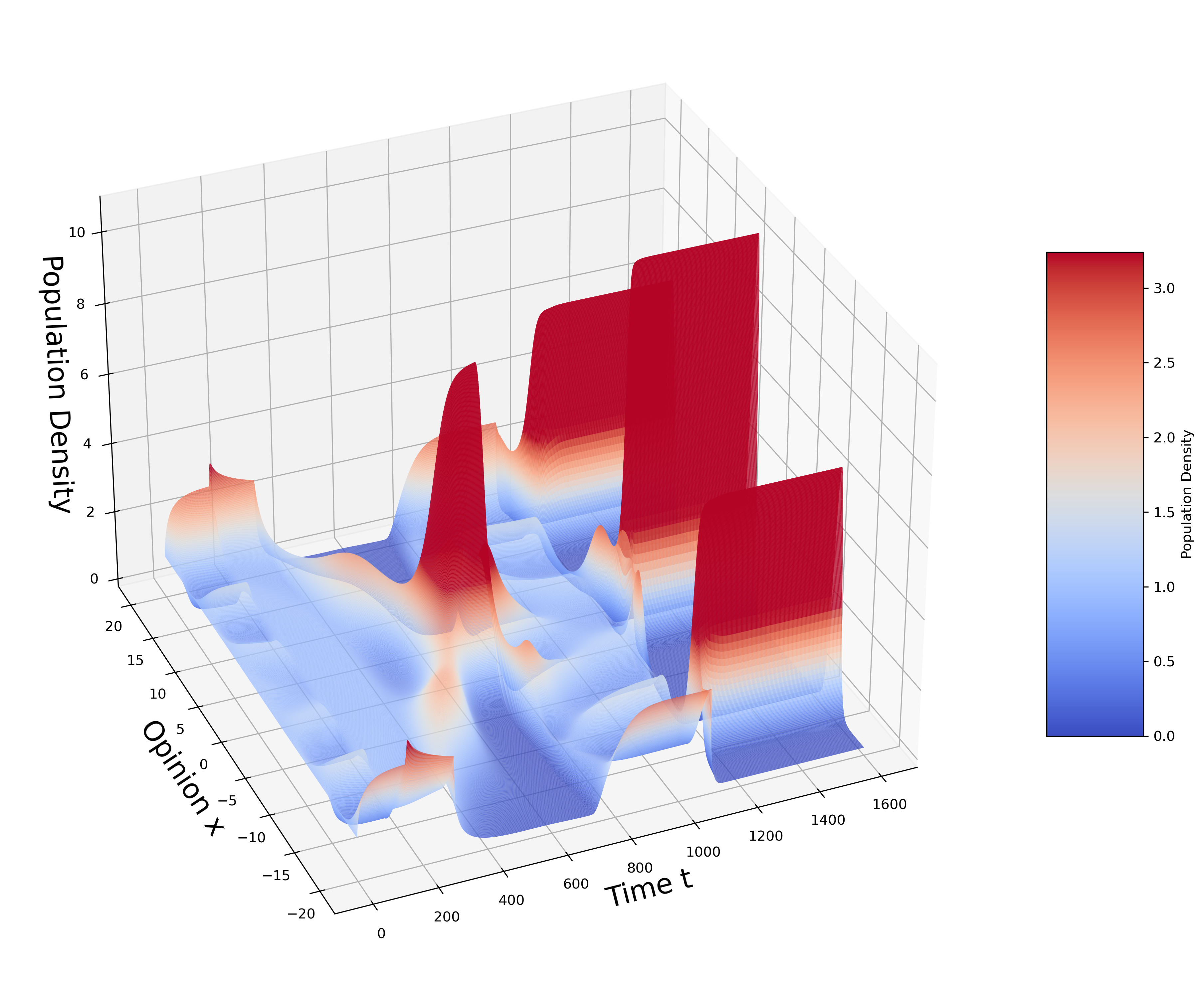}
    \caption{Formation of multiple opinion clusters for Prednisolone control feature without potential field guidance.}
    \label{fig:prednisolone_multi_cluster_formation}
\end{figure}

The final experiment, without utilizing potential fields or target objectives, solely allows agents to update kernel function parameters sigma and mu based on external driving facts variations through the combined methodology of causal effect, spatial heterogeneity, and temporal dynamics. The experimental setup remains identical to Experiment 2, with the only modification being the removal of potential field guidance. As shown in the Figure~\ref{fig:prednisolone_multi_cluster_formation}, the results exhibit wave-like diffusion toward both minimum and maximum values, maintaining this state for a period before multiple small peaks emerge, subsequently forming localized opinion clusters around three stable peaks that constitute collective descriptive norms for Prednisolone. The final population-weighted Wasserstein distance to the GMMs of Prednisolone control feature in period 4 dataset is 8.942314, as illustrated in the Figure~\ref{fig:prednisolone_multi_cluster_formation}.

\section{Discussion of experimental results}
Experiment 1 results demonstrate that, under direct constraints of clinical guidelines and through kernel function updates by comparing individual SINP and target objective distances, our model can converge to form collective descriptive norms in an opinion space with width of 40, achieving a population-weighted Wasserstein distance of 0.2965 to the 3-component projected GMM of 33-dimensional GMMs in actual dataset period 3 on azithromycin control feature. Although the precision of this Wasserstein distance has room for improvement compared to the mean value of 0.008 for the largest component 2 in Figure~\ref{fig:azithromycin_gmm}, considering the scale range of the overall opinion space width of 40, this result remains positive. This indicates that each location agent updates the subjective perception of normal behavior represented by its SINP through kernel function local sampling of surrounding behaviors, thereby achieving the transmission and sharing of guideline-dependent norms at the macro level.

In Experiment 2, the bottom-up experiment no longer possesses the target OBJ serving as a clinical guideline function. As shown in Figure~\ref{fig:prednisolone_evolution_3d_causal_potential}, the net causal effect induces the fitted norm result of auto-avoidance period 3 upon activation. $\mu < 0$ reverses the migration direction of $G(P)$, causing the system to abruptly lose aggregation force; the migration term dominates and rapidly reduces population density in the peak region. At this point, $G(P)$ generates a strong negative gradient at the distribution edges, pushing population away from the center, forming local minima and maxima on both sides (wave-like structure). During the wave diffusion phase, the influence of the potential field $V(x)$ is dominated by nonlocal migration, corresponding to a chaotic period in medical practice where inconsistent physician responses to the new variant form multiple temporary practice patterns (extrema in the wave). As wave amplitude decreases, the potential field’s effect gradually emerges. The system eventually ``settles down''. With wave diffusion, the distribution becomes more dispersed and the magnitude of $G(P)$ diminishes. At this stage, the external potential field becomes the dominant term. Long-term professional training experience still produces collective attraction, pulling the system toward the new target position. As causal effects accumulate, $\mu$ becomes locally positive; time\_amplification amplifies long-term effects, and when the system approaches the new equilibrium point $x_{\mathrm{target}}$, the nonlocal gradient $G(P)$ balances with the potential field gradient $-\nabla V(x)$, forming a new unimodal stable distribution corresponding to a new collective descriptive norm. However, the external spatial potential field has low precision; the final Wasserstein distance to dataset period 4 is 3.4218, indicating neutral accuracy, due to accumulated errors from various approximations in the numerical solution of the dynamic PDE system.

The results of Experiment 3, conducted with fully independent and autonomous interaction, show that the emergence of three significant peaks and a Wasserstein distance of 8.942314 indicate that the collective within the system has not converged to the actual pattern in the dataset's period 4, but rather exhibits multicentric practice fragmentation. This suggests that cumulative causal effects cause $\mu$ to become positive in three specific regions, leading to self-sustaining local norms in each region, while the locality of SINP without guidance prevents the population from perceiving the global optimum.

\section{Conclusion and Future Work}
Threat Rigidity Theory (TRT) \cite{staw1981threat} demonstrates that under threat conditions, attention and communication diminish while decision control tightens: reduced attention manifests as a negative shift in $\mu$ (shifting from exploring diverse dissent to reverting to consensus); reduced communication manifests as a decrease in $\sigma$ (narrower peak width), focusing on core treatment protocols when facing negative outcomes from new virus strains. During crisis periods, learned behavior becomes prominent as individuals struggle to comprehend or critically analyze the situation \cite{weick1988enacted,weick1990vulnerable}, tending to maintain the status quo, simplify decisions, and revert to familiar practices even when inappropriate \cite{beach1978contingency}. This aligns with the parameter changes in the interaction kernel of our mathematical model. 

Our work provides such a baseline, offering a rigorous computable model for studying the violation and convergence of collective descriptive norms. As a real-world dataset example, multiple approaches exist to extend new experiments to other datasets. Current models, including this work, have not yet achieved the ability for agents to induce norms through fully autonomous interaction. Existing AI and autonomous multi-agent systems still require human-predefined learning objectives and training within constrained environments and data~\cite{sayama2015introduction,morris2019norm,li2024survey}. We must expand the opinion space to three dimensions and introduce large language models for prescriptive norm-level testing to achieve norm induction accuracy at the clinical pathway mining level.
\printcredits
\section{Appendix}
\begin{appendices}

\section{Details of linear stability analysis with the potential field}
\label{appendix:Details of linear stability analysis}

After introducing the external potential field \( V(x) = k (x - x_{\text{target}})^2 \), the model equation \ref{eq:1} becomes:
\begin{equation}
\frac{\partial P}{\partial t} = d \nabla^2 P - \nabla \cdot \left[ P \left( c  G(P) - \nabla V(x) \right) \right]
\label{eq:1prime}
\end{equation}

Here, \( \nabla V = 2k (x - x_{\text{target}}) \). To analyze the changes in linear perturbations after introducing the potential field, we similarly replace \( P(x,t) \) with a uniform population level \( P_h \) plus a sinusoidal spatial perturbation~\cite{hillen2009user,di2013measure,sayama2020enhanced} with a time-varying small amplitude \( \Delta P(t) \):
\begin{equation}
P(x,t) \rightarrow P_h + \Delta P(t) \sin(\omega x + \phi).
\label{eq:A1}
\end{equation}

Substituting this into Equation~\eqref{eq:1} yields:
\begin{equation}
\begin{aligned}
\sin(\omega x + \phi) \frac{d \Delta P}{dt} ={} & -d \omega^2 \sin(\omega x + \phi) \Delta P \\
& - c \frac{\partial}{\partial x} [P_h + \Delta P \sin(\omega x + \phi)] \\
& \quad \times \int_{-\infty}^{\infty} \{P_h + \Delta P \sin[\omega(x+y) + \phi]\} g(y)  dy \\
& + \frac{\partial}{\partial x} \left[ (P_h + \Delta P \sin(\omega x + \phi)) \cdot 2k (x - x_{\text{target}}) \right].
\end{aligned}
\label{eq:A2}
\end{equation}

By neglecting the second-order terms of \( \Delta P \) and utilizing the property that \( g(y) \) is an odd function, the linear approximation of the migration term remains:
\begin{equation}
- c P_h \frac{\partial}{\partial x} \int_{-\infty}^{\infty} \Delta P \sin[\omega(x+y) + \phi] g(y)  dy + \text{(potential field term)}.
\label{eq:A3}
\end{equation}

Continuing to simplify the migration term:
\begin{equation}
\begin{aligned}
= -d \omega^2 \sin(\omega x + \phi) \Delta P & + c \omega P_h \sin(\omega x + \phi) \Delta P \\
& \times \int_{-\infty}^{\infty} \sin(\omega y) g(y)  dy \\
& + \text{(potential field term)}.
\end{aligned}
\label{eq:A4}
\end{equation}

Now consider the potential field term:
\begin{equation}
\begin{aligned}
\frac{\partial}{\partial x} & \left[ (P_h + \Delta P \sin(\omega x + \phi)) \cdot 2k (x - x_{\text{target}}) \right] \\
= & 2k P_h + 2k \Delta P \sin(\omega x + \phi) \\
& + 2k (x - x_{\text{target}}) \omega \Delta P \cos(\omega x + \phi).
\end{aligned}
\label{eq:A5}
\end{equation}

\begin{figure}
    \centering
    \includegraphics[width=\textwidth]{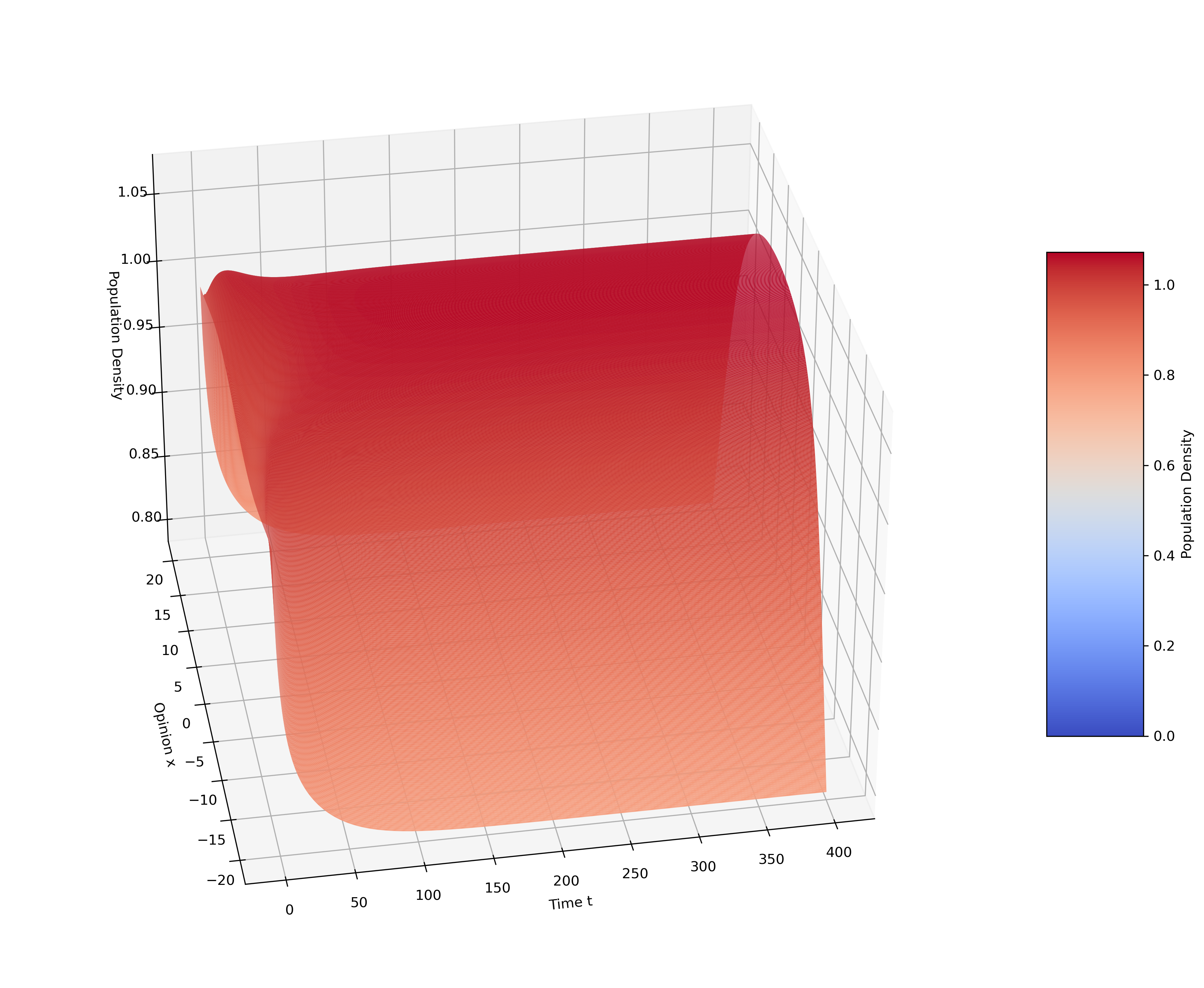}
    \caption{The absence of compression effects.}
    \label{fig:no_potential_field}

\end{figure}

\begin{figure}
    \centering
    \includegraphics[width=\textwidth]{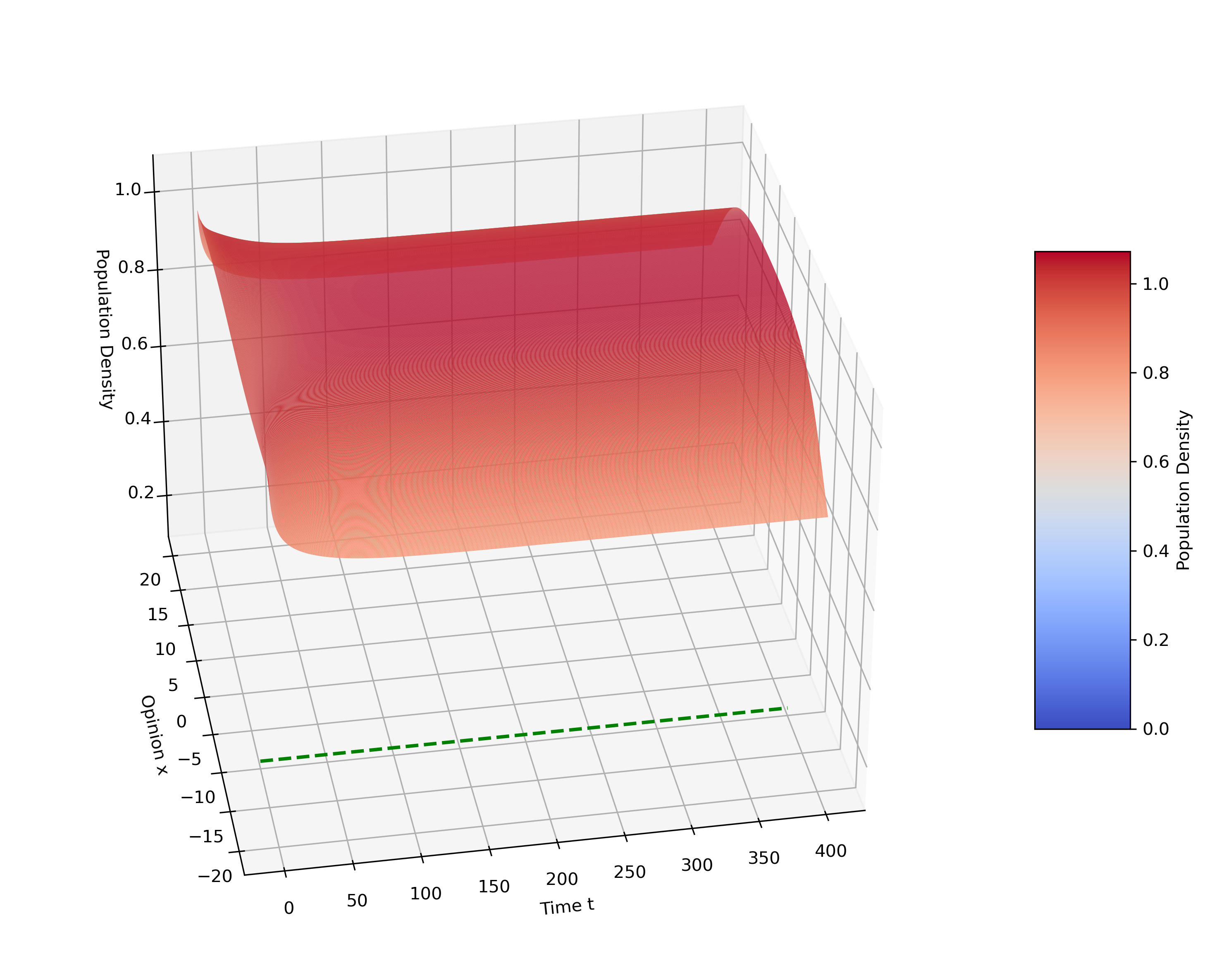}
    \caption{With the target position fixed at $x = -10$, the compression effect attributable to the $2k P_h$ term becomes distinctly observable when compared to the distribution shown in Figure~\ref{fig:no_potential_field}.}
    \label{fig:with_potential_field}

\end{figure}

The net flux induced by the external potential field acting upon the homogeneous background causes time-dependent evolution of the background density. In the linear approximation, the constant term \( 2k P_h \) corresponds to the evolution of the uniform background (attributable to the global compression effect induced by the potential field, as visually demonstrated in Figures~\ref{fig:no_potential_field} and~\ref{fig:with_potential_field} where both cases employ parameter configurations (\( P_h = 1 \),\( c = 1 \),\( k = 0.01 \),\( d = 2 \) and \( w = 0.2 \)) that preclude self-aggregation, isolating purely the temporal evolution of background density). 

For the analysis of perturbation growth rate, we focus on terms linearly related to \( \Delta P \). 
Consider a finite region near \( x_{\text{target}} \), specifically \( [x_{\text{target}} - L, x_{\text{target}} + L] \), where \( L \ll 1/\omega \) (i.e., the perturbation wavelength is much smaller than the region size). Within this region, it can be approximated that \( (x - x_{\text{target}}) \approx 0 \). We approximately neglect the term \( 2k (x - x_{\text{target}}) \omega \Delta P \cos(\omega x + \phi) \). Thus, the linear contribution of the potential field term is approximately \( 2k \Delta P \sin(\omega x + \phi) \).

Combining all terms, dividing by \( \sin(\omega x + \phi) \), and merging the coefficients of \( \Delta P \), we obtain:
\begin{equation}
\begin{aligned}
\frac{d \Delta P}{dt} = \bigg(-d \omega^2 & + c \omega P_h \int_{-\infty}^{\infty} \sin(\omega y) g(y)  dy \\
& + 2k \bigg) \Delta P.
\end{aligned}
\label{eq:A6}
\end{equation}

If the coefficient inside the parentheses is positive, the perturbation grows, indicating destabilization of the uniform distribution. By defining \( Q(\omega) = \int_{-\infty}^{\infty} \frac{\sin(\omega y)}{\omega} g(y)  dy \), the condition for pattern formation is that for \( \omega > 0 \):
\begin{equation}
Q(\omega) > \frac{1}{c P_h} \left(d - \frac{2k}{\omega^2}\right).
\label{eq:A7}
\end{equation}

Compared to the original condition ~\cite{hillen2009user,di2013measure,sayama2020enhanced} \( Q(\omega) > \frac{d}{c P_h} \), after introducing the potential field (assuming \( k > 0 \)), the right-hand side decreases (since \( \frac{2k}{\omega^2} > 0 \)), meaning the instability condition is more easily satisfied. That is, the potential field promotes cluster formation and reduces the critical value of migration strength required. Note that this approximation neglects the spatial variation term of the potential field gradient. If this term is significant, mode coupling must be considered, possibly requiring analysis via numerical methods or higher-order perturbation theory.

We note that $Q(\omega)$ ~\cite{hillen2009user,di2013measure,sayama2020enhanced} is, by itself, the generalized non-local gradient of $\sin(\omega x)/\omega$ around $x=0$. This indicates that the range of $Q(\omega)$ is bounded by the range of the gradients of the original function $\sin(\omega x)/\omega$, which is $\cos(\omega x)$, hence $Q(\omega) \in [-1,1]$.

Moreover, we show that $Q(\omega)$ approaches its maximum 1 regardless of the shape of $g(y)$ in the limit of $\omega \to 0$, as follows:

\begin{equation}
\lim_{\omega \to 0} Q(\omega) = \lim_{\omega \to 0} \int_{-\infty}^{\infty} \frac{\sin(\omega y)}{\omega} g(y) dy 
\label{eq:A9}
\end{equation}

\begin{equation}
= \lim_{\omega \to 0} \int_{-\infty}^{\infty} y \, g(y) dy 
=1\label{eq:A10}
\end{equation}

\begin{figure}
    \centering
    \includegraphics[width=\textwidth]{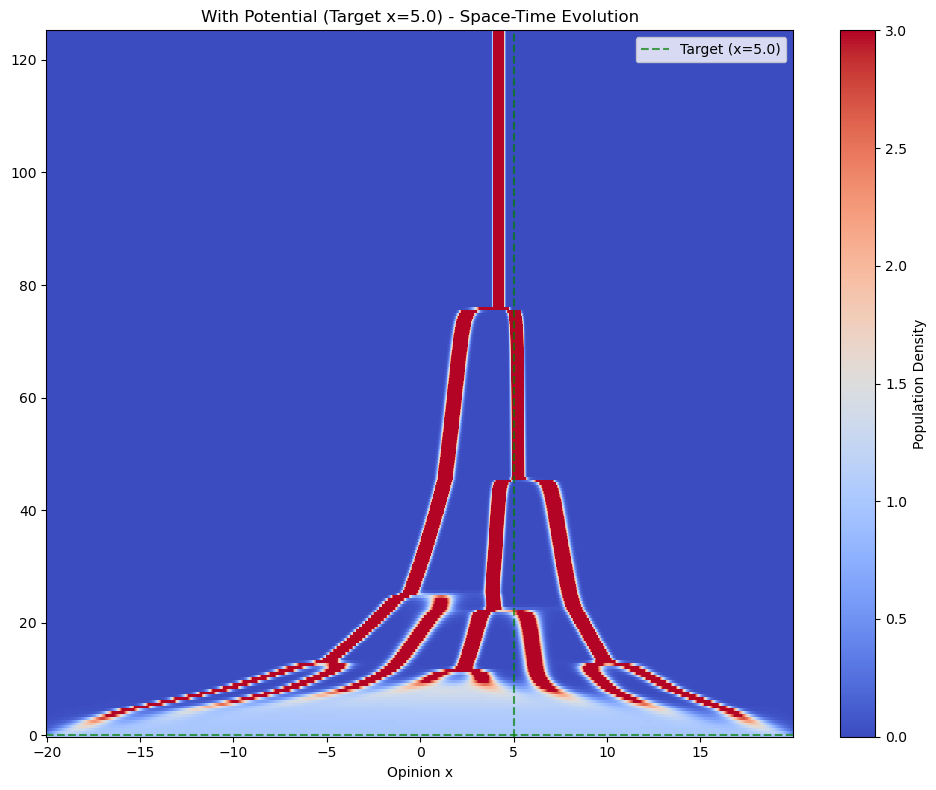} % 替换为实际图像路径
    \caption{Root-like opinion aggregation under baseline parameters: $L=20.0$, $N_x=1000$, $P_0(x)=1.0+0.01\sin(0.2x)$, $d=0.2$, $c=1.0$, $\sigma=1$, $\mu=0.5$, $x_{\mathrm{target}}=5$.}
    \label{fig:baseline_potential_field}
\end{figure}

Under the baseline parameter setting, we fix the domain size as $L=20.0$, spatial resolution as $N_x=1000$, initial density profile as $P_0(x) = 1.0 + 0.01\sin(0.2x)$, diffusion coefficient as $d=0.2$, and migration coefficient as $c=1.0$. We set $\sigma = 1$, $\mu = 0.5$, and $x_{\mathrm{target}} = 5$, which yields a root-like opinion aggregation pattern as shown in Figure~\ref{fig:baseline_potential_field}.

\setcounter{figure}{0}
\setcounter{table}{0}
\renewcommand{\thefigure}{C.\arabic{figure}}
\renewcommand{\thetable}{C.\arabic{table}}

\renewcommand{\theequation}{C.\arabic{equation}}
\setcounter{equation}{0} % Reset counter for appendix

\section{SINP Initialization and Update Formulas}
\label{appendix:SINP}

\subsection{SINP Initialization Formula (at position i)}

Given the initial population distribution $P_0(x)$, spatial grid $x$, and initialization parameters $\sigma_{\text{init}}$ and $\mu_{\text{init}}$, the Subjective Norm Perception (SINP) at position $i$ is initialized as follows:

\begin{align}
&\text{1. Window definition:} \quad \delta = 5 \frac{\max(\sigma_{\text{init}}, |\mu_{\text{init}}|)}{\Delta x}, \\
&\quad\quad\quad\quad\quad\quad\quad\quad\ \mathcal{W}_i = \left[ x_i - \delta,\  x_i + \delta \right] \\
&\text{2. Peak detection:} \quad \mathcal{P}_i = \left\{ p_k \mid \text{find\_peaks}\left(P_0(\mathcal{W}_i)\right) \right\} \\
&\text{3. GMM parameters:} \nonumber \\
&\quad \bullet\ K = \begin{cases} 
1 & \text{if } \mathcal{P}_i = \emptyset \\
|\mathcal{P}_i| & \text{otherwise}
\end{cases} \\
&\quad \bullet\ \pi_k = \frac{P_0(x_{p_k})}{\sum_{j \in \mathcal{P}_i} P_0(x_j)} \quad (\text{normalized peak heights}) \\
&\quad \bullet\ \mu_k = x_{p_k} \quad (\text{peak positions}) \\
&\quad \bullet\ \sigma_k^2 = \max\left( \left( \frac{\text{FWHM}_k}{2} \right)^2,\ 0.001 \right) \\
&\quad\quad \text{where } \text{FWHM}_k = \min_{a,b} \left\{ b - a \mid P_0(x) \geq \frac{1}{2} P_0(x_{p_k})\ \forall x \in [a,b] \right\}
\end{align}

\noindent\textbf{Key notes:}
\begin{itemize}
    \item $\Delta x = x_{i+1} - x_i$ is the spatial step size
    \item $\text{FWHM}_k$ is the \textit{full width at half maximum} of peak $k$
    \item When no peaks are detected ($\mathcal{P}_i = \emptyset$), $x_{p_1} = \arg\max_{x \in \mathcal{W}_i} P_0(x)$
\end{itemize}

\subsection{SINP Update Formula (given data set D)}

Given sampled data points $D = \{d_1, d_2, \dots, d_n\}$ from the local neighborhood, the SINP is updated as follows:

\begin{align}
&\text{1. Kernel density estimation:} \quad \hat{P}(x) = \frac{1}{n h} \sum_{j=1}^n K\left(\frac{x - d_j}{h}\right) \\
&\quad\quad \text{where } K(\cdot) = \mathcal{N}(0,1),\  h = 1.06 \sigma_D n^{-1/5} \\
&\text{2. Peak detection:} \quad \mathcal{P} = \left\{ p_k \mid \text{find\_peaks}\left(\hat{P}(x)\right) \right\} \\
&\text{3. GMM parameters:} \nonumber \\
&\quad \bullet\ K = \begin{cases} 
1 & \text{if } \mathcal{P} = \emptyset \\
|\mathcal{P}| & \text{otherwise}
\end{cases} \\
&\quad \bullet\ \pi_k = \frac{\hat{P}(x_{p_k})}{\sum_{j \in \mathcal{P}} \hat{P}(x_j)} \\
&\quad \bullet\ \mu_k = x_{p_k} \\
&\quad \bullet\ \sigma_k^2 = \max\left( \left( \frac{\text{FWHM}_k}{2} \right)^2,\ 0.001 \right) \\
&\quad\quad \text{where } \text{FWHM}_k = \min_{a,b} \left\{ b - a \mid \hat{P}(x) \geq \frac{1}{2} \hat{P}(x_{p_k})\ \forall x \in [a,b] \right\}
\end{align}

\noindent\textbf{Key notes:}
\begin{itemize}
    \item $\hat{P}(x)$ is the standard Gaussian kernel density estimate
    \item $h$ is the bandwidth (Silverman's rule)
    \item The old model is \textit{completely replaced}: new GMM parameters directly overwrite the old ones (allowing $K$ to change)
    \item When no peaks are detected, $x_{p_1} = \arg\max_x \hat{P}(x)$
\end{itemize}

\subsection{Core Characteristics}

\begin{enumerate}
    \item \textbf{Unified framework:}
    \begin{itemize}
        \item Initialization and update share the same \textit{peak-driven GMM construction logic}
        \item Only difference: data source (initial distribution $P_0$ vs. sampled data $D$)
    \end{itemize}
    
    \item \textbf{Physical interpretation:}
    \begin{itemize}
        \item $\pi_k$: normalized peak height $\rightarrow$ \textit{social influence weight}
        \item $\mu_k$: peak position $\rightarrow$ \textit{norm perception center}
        \item $\sigma_k^2$: $\propto (\text{FWHM})^2$ $\rightarrow$ \textit{norm tolerance} (wider width = higher tolerance)
    \end{itemize}
    
    \item \textbf{Robustness design:}
    \begin{itemize}
        \item $\max(\cdot, 0.001)$ prevents zero variance
        \item Falls back to single-component model when no peaks are detected
        \item FWHM directly relates to the \textit{cognitive ambiguity} of norm perception
    \end{itemize}
\end{enumerate}

\noindent These formulas model subjective norm perception as a \textit{data-driven dynamic probability distribution}. The implementation in code corresponds precisely to the mapping from peaks to GMM parameters in both \texttt{initialize()} and \texttt{update()} methods, with the only difference being the data source.

\setcounter{figure}{0}
\setcounter{table}{0}
\renewcommand{\thefigure}{C.\arabic{figure}}
\renewcommand{\thetable}{C.\arabic{table}}

% Enable proper appendix equation numbering format (B.1, B.2, etc.)
\renewcommand{\theequation}{C.\arabic{equation}}
\setcounter{equation}{0} % Reset counter for appendix

\section{Dataset header}
\label{appendix:Dataset Description}

\begin{table}[ht]
\centering
\caption{Dataset features}
\begin{tabular}{ll}
\toprule
t\_point & Dextran\_stat\_control \\
Omeprazole\_stat\_control & Lisinopril\_stat\_control \\
Nadroparin calcium\_stat\_control & Losartan\_stat\_control \\
Esomeprazole\_stat\_control & Temperature\_dinam\_fact \\
Amlodipine\_stat\_control & Lymphocytes\#\_dinam\_fact \\
Ambroxol\_stat\_control & AST\_dinam\_fact \\
Domperidone\_stat\_control & HR\_dinam\_fact \\
Mebrofenin\_stat\_control & RR\_dinam\_fact \\
Technetium\_stat\_control & Total bilirubin\_dinam\_fact \\
Mometasone\_stat\_control & MPV - Mean platelet volume\_dinam\_fact \\
Bisoprolol\_stat\_control & PCT - Thrombocrit\_dinam\_fact \\
Dexamethasone\_stat\_control & Lymphocytes\%\_dinam\_fact \\
Hydrochlorothiazide\_stat\_control & Decreased consciousness\_dinam\_fact \\
Hydroxychloroquine\_stat\_control & Severity degree by CT\_dinam\_fact \\
Rabeprazole\_stat\_control & Lactate dehydrogenase\_dinam\_fact \\
Enoxaparin sodium\_stat\_control & PDW - Platelet distribution width\_dinam\_fact \\
Perindopril\_stat\_control & age\_stat\_fact \\
Acetylcysteine\_stat\_control & Transfusion\_dinam\_control \\
Azithromycin\_stat\_control & Oxygen therapy\_dinam\_control \\
Valsartan\_stat\_control & NIV\_dinam\_control \\
Methylprednisolone\_stat\_control & MV\_dinam\_control \\
Loratadine\_stat\_control & long\_observation\_tar \\
Chloroquine\_stat\_control & outcome\_tar \\
Sodium chloride\_stat\_control & process\_stages \\
Indapamide\_stat\_control & current\_process\_duration \\
Prednisolone\_stat\_control & admission\_date \\
Atorvastatin\_stat\_control & end\_episode \\
\bottomrule
\end{tabular}
\end{table}

The dataset features are categorized based on their types:

\begin{itemize}
    \item \texttt{\_stat\_control} — Static features describing the control process, do not change over time (in covid\_flow, represented as binary features: 1 indicates the drug is included in the treatment plan, 0 indicates not included)
    \item \texttt{\_dinam\_control} — Dynamic features describing the control process, change over time (in covid\_flow, represented as binary features: 1 indicates the treatment procedure is included in the treatment plan, 0 indicates not included)
    \item \texttt{\_stat\_fact} — Static features describing the controlled process, do not change over time (in covid\_flow, represented as features describing the patient, such as gender, age, and other invariant properties)
    \item \texttt{\_dinam\_fact} — Dynamic features describing the controlled process, change over time (in covid\_flow, represented as indicators describing the patient's state, such as body temperature, percentage of lung infection, etc.)
\end{itemize}

Metadata:
\begin{itemize}
    \item \texttt{t\_point} — Time interval identifier. The entire treatment process is divided into equal-length time periods, \texttt{t\_point} is the time period marker
    \item \texttt{end\_episode} — Treatment process termination identifier: 0 indicates treatment not ended at this \texttt{t\_point}, 1 indicates ended
    \item \texttt{case (index)} — Unique identifier for the observation record (treatment process), e.g., "GACAk+Q"
    \item \texttt{long\_observation\_tar} — Total duration of the treatment process
    \item \texttt{current\_process\_duration} — Current treatment duration (up to \texttt{t\_point})
    \item \texttt{outcome\_tar} — Treatment outcome: 1 indicates death, 0 indicates recovery
\end{itemize}

\section{Tables}
\label{appendix:Tables}

\setcounter{figure}{0}
\setcounter{table}{0}
\renewcommand{\thefigure}{D.\arabic{figure}}
\renewcommand{\thetable}{D.\arabic{table}}
\renewcommand{\theequation}{D.\arabic{equation}}
\setcounter{equation}{0}

\label{appendix:Details of dynamic Medical Factor}

\begin{table}
\centering
\caption{Top 10 medical features by propensity field frequency across periods (descending order)}
\label{tab:top_features}
\setlength{\tabcolsep}{3.5pt}
\begin{tabular}{p{5cm}p{5cm}p{5cm}}
\toprule
\textbf{Period 2} & \textbf{Period 3} & \textbf{Period 4} \\
\midrule
Hydroxychloroquine (static) \hfill 0.0261 & Omeprazole (static) \hfill 0.0238 & Nadroparin Calcium (static) \hfill 0.0229 \\
Omeprazole (static) \hfill 0.0209 & Esomeprazole (static) \hfill 0.0216 & Omeprazole (static) \hfill 0.0201 \\
Ambroxol (static) \hfill 0.0193 & Nadroparin Calcium (static) \hfill 0.0212 & Esomeprazole (static) \hfill 0.0187 \\
Azithromycin (static) \hfill 0.0192 & Dexamethasone (static) \hfill 0.0182 & Amlodipine (static) \hfill 0.0162 \\
Nadroparin Calcium (static) \hfill 0.0187 & Domperidone (static) \hfill 0.0176 & Dexamethasone (static) \hfill 0.0155 \\
Esomeprazole (static) \hfill 0.0182 & Mebrofenin (static) \hfill 0.0176 & Domperidone (static) \hfill 0.0154 \\
Enoxaparin Sodium (static) \hfill 0.0177 & Technetium (static) \hfill 0.0176 & Mebrofenin (static) \hfill 0.0154 \\
Chloroquine (static) \hfill 0.0169 & Mometasone (static) \hfill 0.0176 & Technetium (static) \hfill 0.0154 \\
Amlodipine (static) \hfill 0.0167 & Amlodipine (static) \hfill 0.0173 & Mometasone (static) \hfill 0.0154 \\
Rabeprazole (static) \hfill 0.0148 & Ambroxol (static) \hfill 0.0166 & Prednisolone (static) \hfill 0.0152 \\
\bottomrule
\end{tabular}
\end{table}

\begin{table}
\centering
\caption{Gaussian Mixture Model results for clinical decision patterns across five periods}
\label{tab:gmm_results}
\setlength{\tabcolsep}{1pt}
\renewcommand{\arraystretch}{1}
\begin{tabular}{@{}ccccccc@{}}
\toprule
Period & Mode & Population & Decision & Decision & Mean & Key Treatments \\
 & ID & Proportion (\%) & Sharpness & Coherence & Propensity & \\
\midrule
1 & 1 & 1.01 & 0.883 & 0.949 & 0.059 & esomeprazole, nadroparin calcium, omeprazole \\
1 & 2 & 67.51 & 0.978 & 0.997 & 0.011 & azithromycin, enoxaparin sodium, hydroxychloroquine \\
1 & 3 & 31.48 & 0.975 & 0.996 & 0.012 & enoxaparin sodium, amlodipine, hydroxychloroquine \\
\addlinespace
2 & 1 & 53.97 & 0.979 & 0.995 & 0.010 & azithromycin, ambroxol, hydroxychloroquine \\
2 & 2 & 0.47 & 0.913 & 0.940 & 0.044 & perindopril, hydrochlorothiazide, bisoprolol \\
2 & 3 & 16.43 & 0.969 & 0.993 & 0.016 & hydroxychloroquine, perindopril, amlodipine \\
2 & 4 & 27.95 & 0.971 & 0.993 & 0.014 & mometasone, omeprazole, esomeprazole \\
2 & 5 & 1.17 & 0.893 & 0.944 & 0.054 & domperidone, esomeprazole, omeprazole \\
\addlinespace
3 & 1 & 36.63 & 0.972 & 0.995 & 0.014 & sodium chloride, loratadine, omeprazole \\
3 & 2 & 62.30 & 0.975 & 0.996 & 0.013 & nadroparin calcium, esomeprazole, omeprazole \\
3 & 3 & 1.08 & 0.881 & 0.944 & 0.060 & mometasone, omeprazole, esomeprazole \\
\addlinespace
4 & 1 & 38.11 & 0.979 & 0.996 & 0.011 & dexamethasone, rabeprazole, nadroparin calcium \\
4 & 2 & 37.27 & 0.973 & 0.993 & 0.014 & domperidone, esomeprazole, omeprazole \\
4 & 3 & 24.62 & 0.972 & 0.995 & 0.014 & loratadine, sodium chloride, nadroparin calcium \\
\addlinespace
5 & 1 & 24.18 & 0.978 & 0.994 & 0.011 & domperidone, esomeprazole, omeprazole \\
5 & 2 & 49.02 & 0.977 & 0.996 & 0.012 & hydrochlorothiazide, omeprazole, amlodipine \\
5 & 3 & 26.80 & 0.976 & 0.995 & 0.012 & sodium chloride, dextran, loratadine \\
\bottomrule
\end{tabular}
\footnotesize
\\
\textit{Note:} 
Unique case counts per period: Period 1 (397), Period 2 (426), Period 3 (557), Period 4 (459), Period 5 (153).
\end{table}

\begin{table}
\centering
\caption{Top 10 features with largest rank changes from Period 2 to Period 3}
\label{tab:rank_change_p2_p3}
\setlength{\tabcolsep}{1pt}
\begin{tabular}{p{4.5cm}cccr}
\toprule
\textbf{Feature} & \textbf{P2 Rank} & \textbf{P3 Rank} & \textbf{Change} & \textbf{Frequency Change ($\Delta$)} \\
\midrule
Hydroxychloroquine (static) & 1 & 28 & $\downarrow$27 & $-0.017991$ \\
Azithromycin (static) & 4 & 27 & $\downarrow$23 & $-0.011090$ \\
Chloroquine (static) & 8 & 29 & $\downarrow$21 & $-0.008990$ \\
Dexamethasone (static) & 19 & 4 & $\uparrow$15 & $+0.006909$ \\
Loratadine (static) & 28 & 17 & $\uparrow$11 & $+0.005568$ \\
Sodium Chloride (static) & 29 & 18 & $\uparrow$11 & $+0.004910$ \\
Acetylcysteine (static) & 25 & 15 & $\uparrow$10 & $+0.005380$ \\
Enoxaparin Sodium (static) & 7 & 16 & $\downarrow$9 & $-0.004313$ \\
Dextran (static) & 31 & 23 & $\uparrow$8 & $+0.004617$ \\
Ambroxol (static) & 3 & 10 & $\downarrow$7 & $-0.002677$ \\
\bottomrule
\end{tabular}
\end{table}

\begin{table}
\centering
\caption{Top 10 features with largest rank changes from Period 2 to Period 4}
\label{tab:rank_change_p2_p4}
\begin{tabular}{p{4.5cm}cccr}
\toprule
\textbf{Feature} & \textbf{P2 Rank} & \textbf{P4 Rank} & \textbf{Change} & \textbf{Frequency Change ($\Delta$)} \\
\midrule
Hydroxychloroquine (static) & 1 & 30 & $\downarrow$29 & $-0.018932$ \\
Azithromycin (static) & 4 & 32 & $\downarrow$28 & $-0.012157$ \\
Chloroquine (static) & 8 & 31 & $\downarrow$23 & $-0.009830$ \\
Prednisolone (static) & 32 & 10 & $\uparrow$22 & $+0.008110$ \\
Enoxaparin Sodium (static) & 7 & 25 & $\downarrow$18 & $-0.007805$ \\
Dexamethasone (static) & 19 & 5 & $\uparrow$14 & $+0.004212$ \\
Ambroxol (static) & 3 & 14 & $\downarrow$11 & $-0.005340$ \\
Loratadine (static) & 28 & 18 & $\uparrow$10 & $+0.004506$ \\
Methylprednisolone (static) & 23 & 13 & $\uparrow$10 & $+0.004236$ \\
Dextran (static) & 31 & 22 & $\uparrow$9 & $+0.003853$ \\
\bottomrule
\end{tabular}
\end{table}

\begin{table}
\centering
\caption{Top 10 features with largest rank changes from Period 3 to Period 4}
\label{tab:rank_change_p3_p4}
\begin{tabular}{p{4.5cm}cccr}
\toprule
\textbf{Feature} & \textbf{P3 Rank} & \textbf{P4 Rank} & \textbf{Change} & \textbf{Frequency Change ($\Delta$)} \\
\midrule
Prednisolone (static) & 26 & 10 & $\uparrow$16 & $+0.005212$ \\
Enoxaparin Sodium (static) & 16 & 25 & $\downarrow$9 & $-0.003492$ \\
Methylprednisolone (static) & 19 & 13 & $\uparrow$6 & $+0.001902$ \\
Amlodipine (static) & 9 & 4 & $\uparrow$5 & $-0.001160$ \\
Rabeprazole (static) & 12 & 17 & $\downarrow$5 & $-0.001637$ \\
Azithromycin (static) & 27 & 32 & $\downarrow$5 & $-0.001067$ \\
Transfusion (dynamic) & 33 & 29 & $\uparrow$4 & $+0.000495$ \\
Ambroxol (static) & 10 & 14 & $\downarrow$4 & $-0.002663$ \\
Atorvastatin (static) & 22 & 26 & $\downarrow$4 & $-0.002017$ \\
Oxygen Therapy (dynamic) & 31 & 27 & $\uparrow$4 & $+0.000385$ \\
\bottomrule
\end{tabular}
\end{table}

\begin{table}[ht]
\centering
\caption{Statistical comparison of medical features between Period 1 and Period 2}
\label{tab:period1_2_comparison}
\setlength{\tabcolsep}{3pt} 
\renewcommand{\arraystretch}{2}
\begin{tabular}{p{3cm}ccccccc}
\toprule
Feature & Test Stat. & $p$-value & \begin{tabular}{@{}c@{}}Sample Size\\ {[}P1,P2{]}\end{tabular} & Effect Size ($\delta$) & Direction & $p_{\text{corr}}$ & Sig. (corr) \\
\midrule
Age (static) & 80378.5 & 0.2196 & [397, 426] & -0.049 & Period1 < Period2 & 0.2196 & FALSE \\ 
AST (dynamic) & 82997.5 & 0.54190 & [382, 424] & 0.025 & Period1 < Period2 & 0.70447 & FALSE \\
Total Bilirubin \\ (dynamic) & 82347 & 0.00037 & [359, 399] & 0.150 & Period1 > Period2 & 0.00476 & TRUE \\
LDH (dynamic) & 67904.5 & 0.42054 & [334, 393] & 0.035 & Period1 > Period2 & 0.67430 & FALSE \\
Respiratory Rate \\ (dynamic) & 75547 & 0.46682 & [377, 413] & -0.030 & Period1 < Period2 & 0.67430 & FALSE \\
MPV (dynamic) & 64994.5 & 0.00269 & [365, 407] & -0.125 & Period1 < Period2 & 0.01748 & TRUE \\
Lymphocytes (\%) (dynamic) & 78783.5 & 0.14526 & [365, 407] & 0.061 & Period1 > Period2 & 0.38758 & FALSE \\
CT Severity (dynamic) & 89272.5 & 0.14907 & [397, 426] & 0.056 & Period1 > Period2 & 0.38758 & FALSE \\
PDW (dynamic) & 78223 & 0.20218 & [365, 407] & 0.053 & Period1 > Period2 & 0.43805 & FALSE \\
Temperature (dynamic) & 1232 & 0.90704 & [54, 45] & 0.014 & Period1 < Period2 & 0.95901 & FALSE \\
PCT (dynamic) & 74118 & 0.95901 & [365, 407] & -0.002 & Period1 > Period2 & 0.95901 & FALSE \\
Heart Rate (dynamic) & 80643 & 0.78774 & [392, 416] & -0.011 & Period1 < Period2 & 0.93097 & FALSE \\
Lymphocytes (abs) (dynamic) & 77411.5 & 0.31108 & [365, 407] & 0.042 & Period1 > Period2 & 0.57772 & FALSE \\
Consciousness Impairment (dynamic) & 87610 & 0.03538 & [397, 426] & 0.036 & Period1 < Period2 & 0.15330 & FALSE \\
\bottomrule
\end{tabular}
\end{table} 

\begin{table}
\centering
\caption{Statistical comparison of medical features between Period 3 and Period 4}
\label{tab:period3_4_comparison}
\setlength{\tabcolsep}{2pt} 
\renewcommand{\arraystretch}{1.5}
\begin{tabular}{lccccccc}
\toprule
Feature & Test Stat. & $p$-value & \begin{tabular}{@{}c@{}}Sample Size\\ {[}P3,P4{]}\end{tabular} & Effect Size ($\delta$) & Direction & $p_{\text{corr}}$ & Sig. (corr) \\
\midrule
Age (static) & 113734 & 0.00245 & [557, 459] & -0.110 & Period3 < Period4 & 0.00245 & TRUE \\
AST (dynamic) & 120541 & 0.46492 & [531, 442] & 0.027 & Period3 > Period4 & 0.54946 & FALSE \\
Total Bilirubin \\ (dynamic) & 100978 & 0.00201 & [525, 435] & -0.116 & Period3 < Period4 & 0.00871 & TRUE \\
LDH (dynamic) & 124318 & 0.01064 & [524, 433] & 0.096 & Period3 > Period4 & 0.02766 & TRUE \\
Respiratory Rate \\ (dynamic) & 137498 & 0.00104 & [550, 447] & 0.119 & Period3 < Period4 & 0.00678 & TRUE \\
MPV (dynamic) & 115661.5 & 0.68425 & [525, 434] & 0.015 & Period3 > Period4 & 0.72002 & FALSE \\
Lymphocytes (\%) \\ (dynamic) & 110812.5 & 0.43007 & [525, 435] & -0.030 & Period3 < Period4 & 0.54946 & FALSE \\
CT Severity \\ (dynamic) & 146332.5 & $2.99\times10^{-5}$ & [557, 459] & 0.145 & Period3 > Period4 & 0.00039 & TRUE \\
PDW (dynamic) & 108720 & 0.22281 & [525, 434] & -0.046 & Period3 < Period4 & 0.35236 & FALSE \\
Temperature \\ (dynamic) & 123490 & 0.13411 & [532, 440] & 0.055 & Period3 < Period4 & 0.29058 & FALSE \\
Plateletcrit \\ (dynamic) & 108435.5 & 0.19851 & [525, 434] & -0.048 & Period3 < Period4 & 0.35236 & FALSE \\
Heart Rate \\ (dynamic) & 118941.5 & 0.24394 & [551, 451] & -0.043 & Period3 < Period4 & 0.35236 & FALSE \\
Lymphocytes (abs) \\ (dynamic) & 101943 & 0.00420 & [525, 435] & -0.107 & Period3 < Period4 & 0.01364 & TRUE \\
Consciousness \\ (dynamic) & 128491.5 & 0.72002 & [557, 459] & 0.005 & Period3 < Period4 & 0.72002 & FALSE \\
\bottomrule
\end{tabular}
\end{table}

\begin{table}
\centering
\caption{Statistical comparison of medical features between Period 2 and Period 3}
\label{tab:period2_3_comparison}
\setlength{\tabcolsep}{2pt}
\renewcommand{\arraystretch}{2}
\begin{tabular}{p{2cm}ccccccc}
\toprule
Feature & Test Stat. & $p$-value & \begin{tabular}{@{}c@{}}Sample Size\\ {[}P2,P3{]}\end{tabular} & Effect Size ($\delta$) & Direction & $p_{\text{corr}}$ & Sig. (corr) \\
\midrule
Age (static) & 113160.5 & 0.2140 & [426, 557] & -0.046 & Period2 < Period3 & 0.2140 & FALSE \\
Temperature (dynamic) & 13858 & 0.0759 & [45, 532] & 0.158 & Period2 > Period3 & 0.0897 & FALSE \\
Total Bilirubin (dynamic) & 129046.5 & $1.45\times10^{-9}$ & [399, 525] & 0.232 & Period2 > Period3 & $3.78\times10^{-9}$ & TRUE \\
Respiratory Rate (dynamic) & 102079 & 0.0065 & [413, 550] & -0.101 & Period2 < Period3 & 0.0093 & TRUE \\
Lymphocytes (\%) (dynamic) & 150856.5 & $3.46\times10^{-27}$ & [407, 525] & 0.412 & Period2 > Period3 & $1.50\times10^{-26}$ & TRUE \\
MPV (dynamic) & 60009.5 & $1.49\times10^{-30}$ & [407, 525] & -0.438 & Period2 < Period3 & $9.65\times10^{-30}$ & TRUE \\
Heart Rate (dynamic) & 139433 & $7.70\times10^{-9}$ & [416, 551] & 0.217 & Period2 > Period3 & $1.67\times10^{-8}$ & TRUE \\
CT Severity (dynamic) & 109025 & 0.0221 & [426, 557] & -0.081 & Period2  < Period3 & 0.0287 & TRUE \\
Lymphocytes (abs) (dynamic) & 100083.5 & 0.0975 & [407, 525] & -0.063 & Period2 < Period3 & 0.1057 & FALSE \\
Plateletcrit (dynamic) & 91496.5 & 0.0002 & [407, 525] & -0.144 & Period2 < Period3 & 0.0003 & TRUE \\
AST (dynamic) & 138294.5 & $1.25\times10^{-9}$ & [424, 531] & 0.228 & Period2 > Period3 & $3.78\times10^{-9}$ & TRUE \\
Consciousness Level (dynamic) & 117395 & 0.4664 & [426, 557] & -0.011 & Period2 < Period3 & 0.4664 & FALSE \\
LDH (dynamic) & 81564.5 & $6.97\times10^{-8}$ & [393, 524] & -0.208 & Period2 < Period3 & $1.29\times10^{-7}$ & TRUE \\
PDW (dynamic) & 179845 & $9.48\times10^{-72}$ & [407, 525] & 0.683 & Period2 > Period3 & $1.23\times10^{-70}$ & TRUE \\
\bottomrule
\end{tabular}
\end{table}

\begin{sidewaystable}
\centering
\small
\caption{The causal statistical impact of medical facts on medical control measures}
\label{tab:facts_effec_control}
\setlength{\tabcolsep}{2pt}
\begin{tabular}{@{}llcccccc@{}}
\toprule
Control Variable & Driving Fact & \begin{tabular}{c}Causal\\Effect\end{tabular} & \begin{tabular}{c}Standard\\Error\end{tabular} & $p$-value & \begin{tabular}{c}Corrected\\$p$-value\end{tabular} & 95\% CI & Sig. \\
\midrule
omeprazole\_stat\_control & Lymphocytes\#\_dinam\_fact & 0.00224 & 0.000315 & $1.19 \times 10^{-12}$ & $2.38 \times 10^{-11}$ & [0.00162, 0.00286] & *** \\
omeprazole\_stat\_control & long\_observation\_tar & 0.00332 & 0.000564 & $4.15 \times 10^{-9}$ & $4.15 \times 10^{-8}$ & [0.00221, 0.00442] & *** \\
omeprazole\_stat\_control & current\_process\_duration & 0.00205 & 0.000691 & 0.00306 & 0.01531 & [0.000692, 0.00340] & * \\
omeprazole\_stat\_control & days\_since\_pandemic\_start & 0.00379 & 0.000852 & $8.48 \times 10^{-6}$ & $5.65 \times 10^{-5}$ & [0.00212, 0.00546] & *** \\
nadroparin calcium\_stat\_control & Temperature\_dinam\_fact & 0.00357 & 0.00104 & 0.000643 & 0.00214 & [0.00152, 0.00561] & ** \\
nadroparin calcium\_stat\_control & Lymphocytes\#\_dinam\_fact & 0.00269 & 0.000343 & $4.44 \times 10^{-15}$ & $2.96 \times 10^{-14}$ & [0.00201, 0.00336] & *** \\
nadroparin calcium\_stat\_control & consciousness\_decrease\_dinam\_fact & -0.0549 & 0.0184 & 0.00287 & 0.00821 & [-0.0909, -0.0188] & ** \\
nadroparin calcium\_stat\_control & Lactate dehydrogenase\_dinam\_fact & -0.0000975 & 0.0000363 & 0.00724 & 0.01679 & [-0.000169, -0.0000264] & * \\
nadroparin calcium\_stat\_control & PDW\_dinam\_fact & 0.00421 & 0.00158 & 0.00755 & 0.01679 & [0.00112, 0.00730] & * \\
nadroparin calcium\_stat\_control & long\_observation\_tar & 0.00496 & 0.000552 & $<10^{-15}$ & $<10^{-15}$ & [0.00388, 0.00604] & *** \\
nadroparin calcium\_stat\_control & outcome\_tar & -0.124 & 0.0354 & 0.000458 & 0.00183 & [-0.193, -0.0546] & ** \\
nadroparin calcium\_stat\_control & current\_process\_duration & 0.00325 & 0.000652 & $6.07 \times 10^{-7}$ & $3.03 \times 10^{-6}$ & [0.00197, 0.00453] & *** \\
nadroparin calcium\_stat\_control & end\_epizode & -0.0363 & 0.0149 & 0.0149 & 0.0299 & [-0.0655, -0.00707] & * \\
nadroparin calcium\_stat\_control & days\_since\_pandemic\_start & 0.00972 & 0.000805 & $<10^{-15}$ & $<10^{-15}$ & [0.00815, 0.0113] & *** \\
esomeprazole\_stat\_control & Lymphocytes\#\_dinam\_fact & 0.00324 & 0.000375 & $<10^{-15}$ & $<10^{-15}$ & [0.00251, 0.00397] & *** \\
esomeprazole\_stat\_control & days\_since\_pandemic\_start & 0.00368 & 0.000982 & 0.000183 & 0.00183 & [0.00175, 0.00560] & ** \\
amlodipine\_stat\_control & Temperature\_dinam\_fact & -0.00292 & 0.00108 & 0.00707 & 0.02021 & [-0.00504, -0.000794] & * \\
amlodipine\_stat\_control & MPV\_dinam\_fact & 0.00628 & 0.00257 & 0.01457 & 0.03237 & [0.00124, 0.01131] & * \\
\bottomrule
\end{tabular}
\end{sidewaystable}

\end{appendices}

\clearpage
\bibliographystyle{model1-num-names}
% \bibliographystyle{cas-model2-names}

\bibliography{cas-refs}


\end{document}